\documentclass[pra,reprint,aps,superscriptaddress,showpacs,showkeys,amsmath,amssymb]{revtex4-2}
\usepackage{graphicx,xcolor,dcolumn,bm,hyperref,times,mathtools}
\usepackage[T1]{fontenc}
\definecolor{linkblue}{rgb}{0, 0, 1}
\hypersetup{
	colorlinks=true,
	linkcolor=linkblue,
	citecolor=linkblue,
	urlcolor=linkblue,
}
\begin{document}

\title{Nonlinear-linear duality for multipath quantum interference}

\author{Yi~Zheng}
\affiliation{Laboratory of Quantum Information, University of Science and Technology of China, Hefei 230026, China}
\affiliation{Anhui Province Key Laboratory of Quantum Network, University of Science and Technology of China, Hefei 230026, China}
\affiliation{CAS Center for Excellence in Quantum Information and Quantum Physics, University of Science and Technology of China, Hefei 230026, China}

\author{Jin-Shi~Xu}
\email{jsxu@ustc.edu.cn}
\affiliation{Laboratory of Quantum Information, University of Science and Technology of China, Hefei 230026, China}
\affiliation{Anhui Province Key Laboratory of Quantum Network, University of Science and Technology of China, Hefei 230026, China}
\affiliation{CAS Center for Excellence in Quantum Information and Quantum Physics, University of Science and Technology of China, Hefei 230026, China}
\affiliation{Hefei National Laboratory, University of Science and Technology of China, Hefei 230088, China}

\author{Chuan-Feng~Li}
\email{cfli@ustc.edu.cn}
\affiliation{Laboratory of Quantum Information, University of Science and Technology of China, Hefei 230026, China}
\affiliation{Anhui Province Key Laboratory of Quantum Network, University of Science and Technology of China, Hefei 230026, China}
\affiliation{CAS Center for Excellence in Quantum Information and Quantum Physics, University of Science and Technology of China, Hefei 230026, China}
\affiliation{Hefei National Laboratory, University of Science and Technology of China, Hefei 230088, China}

\author{Guang-Can~Guo}
\affiliation{Laboratory of Quantum Information, University of Science and Technology of China, Hefei 230026, China}
\affiliation{Anhui Province Key Laboratory of Quantum Network, University of Science and Technology of China, Hefei 230026, China}
\affiliation{CAS Center for Excellence in Quantum Information and Quantum Physics, University of Science and Technology of China, Hefei 230026, China}
\affiliation{Hefei National Laboratory, University of Science and Technology of China, Hefei 230088, China}

\date{\today}

\begin{abstract}
In quantum optics, the postselection amplitude of a nondegenerate parametric down-conversion (PDC) process is linked to a beamsplitter (BS) via partial time reversal, up to a normalization coefficient which is related to the parametric gain [\href{https://doi.org/10.1073/pnas.2010827117}{Proc.\ Natl.\ Acad.\ Sci.\ USA {\bf 117}, 33107 (2020)}]. A special example where the gain is low is reminiscent of Klyshko's advanced-wave picture in quantum imaging. Here, we propose and prove a generalized duality for multiple spatial paths connecting a quantum nonlinear interference setup consisting of nondegenerate PDCs and linear optical systems to a linear one, where the PDCs are directly replaced by hypothetical wavelength-shifting BSs. This replacement preserves the geometry of the original setup, and cascaded PDCs become optical cavities whose calculation involves the Redheffer star product. Additional terms in the normalization coefficient are related to the contribution of looping photons inside the cavities. Then, we discuss the case of coherent state input and postselection for \emph{Q}-function calculation. This theorem will be helpful in the development of quantum photonic devices beyond the low-gain limit.
\end{abstract}

\maketitle

\section{Introduction}

Nonclassical light states play a central role in quantum optics and quantum information technology \cite{DELLANNO200653,RevModPhys.92.035005,Quesada:22,Zhang:24,quanphoton}. Quantum interference phenomena \cite{Chekhova:16,JIN2024100519}, including the Hong-Ou-Mandel (HOM) effect \cite{PhysRevLett.59.2044,Bouchard_2021,PhysRevResearch.3.043065}, boson sampling \cite{science.1231692,Tillmann2013} with its variants \cite{PhysRevLett.113.100502,PhysRevLett.119.170501,PhysRevA.100.032326}, and the Zou-Wang-Mandel effect \cite{PhysRevLett.67.318,PhysRevA.44.4614} with its further researches on path identity \cite{PhysRevLett.119.240403,RevModPhys.94.025007,Qian2023} and imaging \cite{Lemos2014,PhysRevA.92.013832,BarretoLemos:22}, are described by the transformation of creation and annihilation operators in the multimode Fock space \cite{Leonhardt_2003}. For two spatial modes denoted by $\hat{a}^\dagger_1$ and $\hat{a}^\dagger_2$, a beamsplitter (BS) coupling them is described by the quantum operator $\exp(\theta\hat{a}^\dagger_1\hat{a}_2-\theta^\ast\hat{a}_1\hat{a}^\dagger_2)$, and a parametric down-conversion (PDC) process, also known as parametric amplification or two-mode squeezing, is $\exp(r\hat{a}^\dagger_1\hat{a}^\dagger_2-r^\ast\hat{a}_1\hat{a}_2)$, where $r$ is proportional to the nonlinear coefficient and the pump light amplitude. The similarity in their expressions suggests a possible connection between them.

In 2020, Cerf and Jabbour proposed the nonlinear-linear duality which connects the postselection amplitude of a PDC with given initial and final states to that of a BS by partial time reversal (PTR) \cite{pnas.2010827117}. In classical optics, for a monochromatic light field passing through a linear, passive optical system, time reversal is related to the optical reciprocity \cite{carsonrt,PhysRevA.110.063710}, i.e., describing the system as a scattering matrix, the backward system (from the output to the input) is the transpose of the original matrix (in this article, we let all the transpose and complex conjugate operations act in the multimode Fock basis). For a PDC process, if the idler (\emph{i}) light is time-reversed while the signal (\emph{s}) one is not, the resulting operator is the partial transpose of the original one with respect to the \emph{i} modes. They proved the partial transpose of the PDC operator is the BS operator times a normalization coefficient (NC) (Salazar \emph{et al.}\ interpreted it from the perspective of group theory \cite{grouptheory}). This BS shifts the wavelength when reflecting and is, thus, hypothetical \cite{twm}, but its action can still be treated as a linear optical process. Then, they focused on an HOM-like effect for PDCs with specific gain values, which was recently experimentally demonstrated by Chen \emph{et al.}\ \cite{chen2025}, and mentioned that this duality is reminiscent of Klyshko's advanced-wave picture (AWP) \cite{Klyshko_1988,KLYSHKO1988133} which is mainly used in quantum ghost imaging \cite{ghostimaging,quanphoton}. For a single PDC, the widely-known AWP is that, if the initial state of the PDC is vacuum (spontaneous PDC; SPDC) and the created photons pass through linear optical systems, considering the two-photon term, the final state of the \emph{i} light becomes the advanced-wave source propagating backward, a thin nonlinear crystal performing SPDC becomes a mirror which creates the retarded \emph{s} wave traveling forward, and the two-photon joint amplitude becomes the \emph{s} light amplitude in this classical model.

\begin{figure*}[t]
\includegraphics[width=0.92\textwidth]{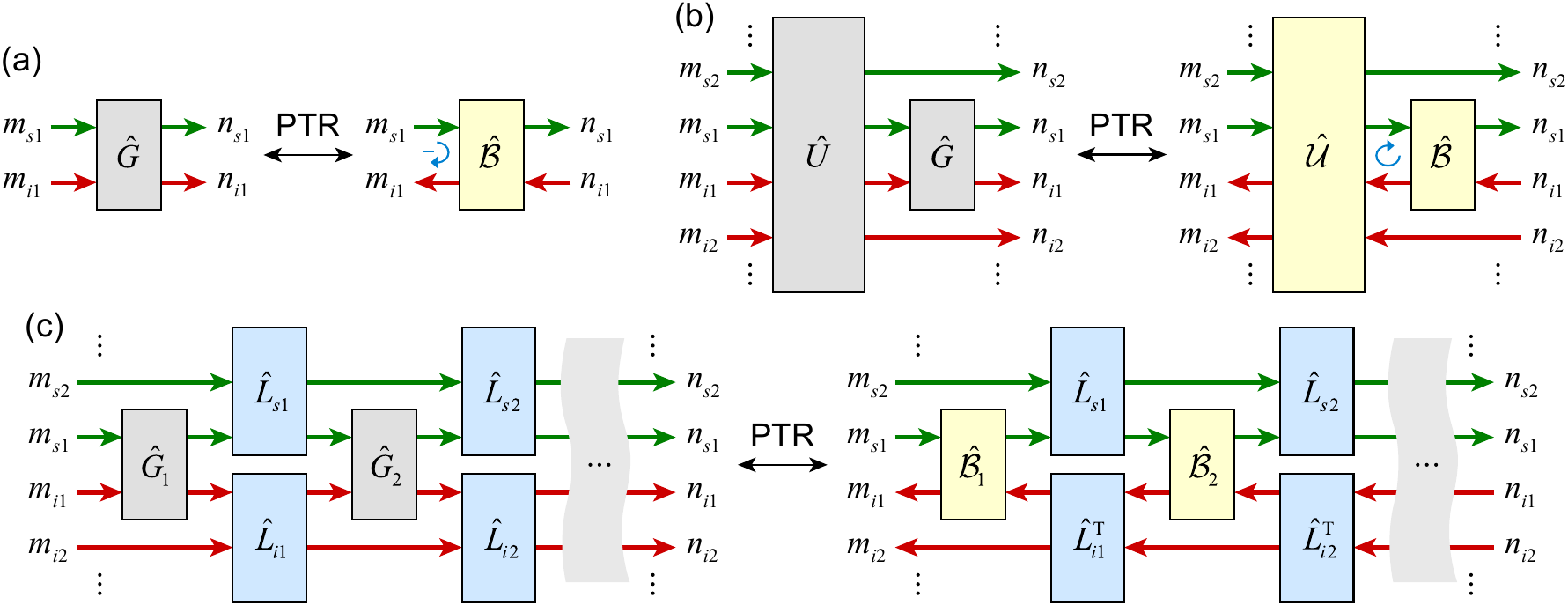}
\caption{\label{ptrfig}(a) The nonlinear-linear duality for a single device proposed by Cerf and Jabbour. The $m$ and $n$ symbols stand for the initial and postselected photon numbers respectively. Green and red arrows represent \emph{s} and \emph{i} lights. The reflected light from $s1$ to $i1$ at the hypothetical BS $\hat{\mathcal{B}}$ gains a phase $\pi$. (b) If a nonlinear system $\hat{U}$ satisfies the duality which links it to a linear setup $\hat{\mathcal{U}}$ which contains hypothetical BSs, a combination of it and a PDC $\hat{G}$ also does. A cavity forms in its partially time-reversed (PTR) setup. (c) The decomposition of a general nonlinear system with nondegenerate PDCs $\hat{G}_n$ and linear systems $\hat{L}_{sn},\hat{L}_{in}$ and its PTR setup.}
\end{figure*}
Recently, we developed a comprehensive AWP theory for monochromatic \emph{s} and \emph{i} lights from an arbitrary optical system consisting of low-gain SPDC crystals of any shape and linear, passive media \cite{PhysRevA.110.063710}. Low-gain SPDC means $r\ll1$, only the first-order (two-photon) term of the SPDC is significant \cite{WALBORN201087}, the corresponding reflectance values of the BSs replacing those nonlinear crystals are low, and the transmitted light is unchanged. In this work, we propose the nonlinear-linear duality for multiple spatial paths, without the low-gain and spontaneous requirement. The main step is to prove the combination of a nonlinear optical system which satisfies the duality and another PDC also does. The new PTR setup generally forms a cavity, and the additional term to the NC is explained. Finally, we discuss two simple examples and the possible applications of this duality including the calculation of the Husimi \emph{Q}-function after a nonlinear system. In the appendixes, we provide another proof of this duality using the Gaussian state theory and a four-photon AWP for nondegenerate SPDC which is purely classical.

\section{Theory}

There are multiple paths for the \emph{s} and \emph{i} lights, $s1,s2,\ldots$ and $i1,i2,\ldots$, and a multimode Fock state is $|\mathbf{n}_s;\mathbf{n}_i\rangle$, where $\mathbf{n}_s=(n_{s1},n_{s2},\ldots)$ and $\mathbf{n}_i=(n_{i1},n_{i2},\ldots)$. For a PDC acting on $s1$ and $i1$, the operator is $\hat{G}=\exp[r(\hat{a}^\dagger_{s1}\hat{a}^\dagger_{i1}-\hat{a}_{s1}\hat{a}_{i1})]$ ($r>0$; the case of a complex squeezing parameter can be realized by adding phase plates). For a hypothetical BS $\hat{\mathcal{B}}=\exp[\theta(\hat{a}^\dagger_{s1}\hat{a}_{i1}-\hat{a}_{s1}\hat{a}^\dagger_{i1})]$ ($0<\theta<\pi/2$; we use calligraphic symbols to denote linear setups with hypothetical BSs), denoting its amplitude transmittance $T=\cos\theta$ and reflectance $R=\sin\theta$, we have
\begin{equation}
    \hat{\mathcal{B}}\hat{a}^\dagger_{s1}\hat{\mathcal{B}}^\dagger=T\hat{a}^\dagger_{s1}-R\hat{a}^\dagger_{i1},\;\hat{\mathcal{B}}\hat{a}^\dagger_{i1}\hat{\mathcal{B}}^\dagger=R\hat{a}^\dagger_{s1}+T\hat{a}^\dagger_{i1},
\end{equation}
which means the reflected light from \emph{s} to \emph{i} gains a $\pi$ phase. Denoting
\begin{gather}
    (\prescript{m_{s1}}{m_{i1}}{}|\hat{G}|^{n_{s1}}_{n_{i1}})=\langle n_{s1};n_{i1}|\hat{G}|m_{s1};m_{i1}\rangle,\nonumber\\
    \{\prescript{m_{s1}}{m_{i1}}{}|\hat{\mathcal{B}}|^{n_{s1}}_{n_{i1}}\}=\langle n_{s1};m_{i1}|\hat{\mathcal{B}}|m_{s1};n_{i1}\rangle
\end{gather}
for simplicity (notice the resemblance between the two symbols and Fig.\ \ref{ptrfig}), Cerf and Jabbour's duality is
\begin{equation}
    (\prescript{m_{s1}}{m_{i1}}{}|\hat{G}|^{n_{s1}}_{n_{i1}})=T\{\prescript{m_{s1}}{m_{i1}}{}|\hat{\mathcal{B}}|^{n_{s1}}_{n_{i1}}\}
\end{equation}
with $\sin\theta=\tanh r$ and the NC $T=\operatorname{sech}r$, as shown in Fig.\ \ref{ptrfig}(a). It is also written as $\hat{G}^{\mathrm{T}_i}=T\hat{\mathcal{B}}$ in the form of partial transpose \cite{pnas.2010827117}. Also, $\{\prescript{0}{0}{}|\hat{\mathcal{B}}|^{0}_{0}\}=1$, so the NC equals $(\prescript{0}{0}{}|\hat{G}|^{0}_{0})$.

Given a linear lossless optical system $\hat{L}$ ($\hat{L}|\mathrm{vac}\rangle=|\mathrm{vac}\rangle$, where $|\mathrm{vac}\rangle$ is the vacuum state) for a certain wavelength, denoting its classical amplitude transfer as the unitary scattering matrix $\boldsymbol{L}$ (an arbitrary unitary $\boldsymbol{L}$ can be realized by BSs and phase plates \cite{PhysRevLett.73.58}), one of its elements $L_{kj}$ ($|L_{kj}|\leq1$) is the output amplitude at path $k$ from the input at path $j$ of unit amplitude, and we have the transformations of creation and annihilation operators \cite{Leonhardt_2003}
\begin{equation}
    \hat{L}\hat{a}_j^\dagger\hat{L}^\dagger=\sum_kL_{kj}\hat{a}_k^\dagger,\;
    \hat{L}^\dagger\hat{a}_j\hat{L}=\sum_kL_{jk}\hat{a}_k.
\end{equation}
$\hat{a}^\mathrm{T}=\hat{a}^\dagger$ because its matrix elements in the Fock basis are real \cite{pnas.2010827117}, so
\begin{equation}
    \hat{L}^\mathrm{T}\hat{a}_j^\dagger(\hat{L}^\mathrm{T})^\dagger
    =(\hat{L}^\dagger\hat{a}_j\hat{L})^\mathrm{T}=\sum_kL_{jk}\hat{a}_k^\dagger=\sum_kL^\mathrm{T}_{kj}\hat{a}_k^\dagger.
\end{equation}
$\boldsymbol{L}^\mathrm{T}$ describes the backward system of $\boldsymbol{L}$, so $\hat{L}^\mathrm{T}$ is its corresponding quantum operator \cite{PhysRevA.98.062314}.

Letting the initial and postselected state of a quantum nonlinear optical system be $|\mathbf{m}_s;\mathbf{m}_i\rangle$ and $|\mathbf{n}_s;\mathbf{n}_i\rangle$ respectively, denoting $\|\mathbf{n}\|=\sum_jn_j$, the \emph{s} and \emph{i} photon number difference should be unchanged after it $\|\mathbf{n}_s\|-\|\mathbf{n}_i\|=\|\mathbf{m}_s\|-\|\mathbf{m}_i\|$. If a nonlinear system $\hat{U}$ satisfies the duality with its PTR linear system $\hat{\mathcal{U}}$, i.e.,
\begin{equation}
    (\prescript{\mathbf{m}_s}{\mathbf{m}_i}{}|\hat{U}|^{\mathbf{n}_s}_{\mathbf{n}_i})=C_U\{\prescript{\mathbf{m}_s}{\mathbf{m}_i}{}|\hat{\mathcal{U}}|^{\mathbf{n}_s}_{\mathbf{n}_i}\},
\end{equation}
letting $\hat{L}_s$ and $\hat{L}_i$ be linear lossless optical systems acting on the \emph{s} and \emph{i} light respectively, we have
\begin{align}
    &\,(\prescript{\mathbf{m}_s}{\mathbf{m}_i}{}|\hat{L}_s\hat{L}_i\hat{U}|^{\mathbf{n}_s}_{\mathbf{n}_i})\nonumber\\
    =&\sum_{\mathbf{n}'_s,\mathbf{n}'_i}\langle\mathbf{n}_s|\hat{L}_s|\mathbf{n}'_s\rangle\langle\mathbf{n}_i|\hat{L}_i|\mathbf{n}'_i\rangle\langle\mathbf{n}'_s;\mathbf{n}'_i|\hat{U}|\mathbf{m}_s;\mathbf{m}_i\rangle\nonumber\\
    =&\,C_U\sum_{\mathbf{n}'_s,\mathbf{n}'_i}\langle\mathbf{n}_s|\hat{L}_s|\mathbf{n}'_s\rangle\langle\mathbf{n}'_s;\mathbf{m}_i|\hat{\mathcal{U}}|\mathbf{m}_s;\mathbf{n}'_i\rangle\langle\mathbf{n}'_i|\hat{L}_i^\mathrm{T}|\mathbf{n}_i\rangle\nonumber\\
    =&\,C_U\{\prescript{\mathbf{m}_s}{\mathbf{m}_i}{}|\hat{L}_s\hat{\mathcal{U}}\hat{L}_i^\mathrm{T}|^{\mathbf{n}_s}_{\mathbf{n}_i}\}.
\end{align}
Currently, we only know $\hat{G}$ satisfies the duality, so the PTR setup of a PDC followed by linear systems is $\hat{\mathcal{B}}$ followed by $\hat{L}_s$ and the backward system of $\hat{L}_i$, with the NC unchanged. Then, given a certain nonlinear system satisfying the duality which is described by $\hat{U}$, $\hat{\mathcal{U}}$, and $C_U$, if we can prove the new system after adding a PDC $\hat{U}'=\hat{G}\hat{U}$ also does [as shown in Fig.\ \ref{ptrfig}(b); possibly with a new NC] and find the corresponding PTR setup, our final result can be proved by mathematical induction, as an arbitrary system with nondegenerate PDCs and linear systems can be decomposed into these steps: a PDC acting on $s1$ and $i1$, two linear systems for \emph{s} and \emph{i} lights respectively, another PDC acting on $s1$ and $i1$, two other linear systems, etc. If a PDC actually acts on other paths, path swaps can be added to the linear systems.

Denoting $\mathbf{n}_{s+}=(n_{s2},n_{s3},\ldots)$, $\mathbf{n}_{i+}=(n_{i2},n_{i3},\ldots)$, and $\Delta=n_{s1}-n_{i1}$, we assume $\Delta\geq0$ without loss of generality (otherwise, the labels \emph{s} and \emph{i} can be swapped). By inserting a complete basis, noting that $\hat{G}$ does not act on paths other than $s1$ and $i1$, we have
\begin{align}\label{insertbasisptr}
    (\prescript{\mathbf{m}_s}{\mathbf{m}_i}{}|\hat{U}'|^{\mathbf{n}_s}_{\mathbf{n}_i})&=\sum_{n'=0}^{+\infty}(\prescript{\mathbf{m}_s}{\mathbf{m}_i}{}|\hat{U}|^{\Delta+n',\mathbf{n}_{s+}}_{n',\mathbf{n}_{i+}})(\prescript{\Delta+n'}{n'}{}|\hat{G}|^{n_{s1}}_{n_{i1}})\nonumber\\
    &=C_UT\sum_{n'=0}^{+\infty}\{\prescript{\mathbf{m}_s}{\mathbf{m}_i}{}|\hat{\mathcal{U}}|^{\Delta+n',\mathbf{n}_{s+}}_{n',\mathbf{n}_{i+}}\}\{\prescript{\Delta+n'}{n'}{}|\hat{\mathcal{B}}|^{n_{s1}}_{n_{i1}}\}.
\end{align}
The expression means the optical states between the two systems can be everything as long as the probability amplitude is nonzero, which might be related to an optical cavity.

We classically examine the combined cavity system $\hat{\mathcal{U}'}$ from $\hat{\mathcal{U}}$ and $\hat{\mathcal{B}}$. With the unitary scattering matrices describing $\hat{\mathcal{U}}$ and $\hat{\mathcal{B}}$ respectively,
\begin{equation}
    \boldsymbol{\mathcal{U}}=\begin{pmatrix}\boldsymbol{\mathcal{U}}_{ss}&\boldsymbol{\mathcal{U}}_{si}\\\boldsymbol{\mathcal{U}}_{is}&\boldsymbol{\mathcal{U}}_{ii}\end{pmatrix},\;
    \boldsymbol{\mathcal{B}}=\begin{pmatrix}\boldsymbol{\mathcal{B}}_{ss}&\boldsymbol{\mathcal{B}}_{si}\\\boldsymbol{\mathcal{B}}_{is}&\boldsymbol{\mathcal{B}}_{ii}\end{pmatrix},
\end{equation}
where
\begin{gather}
    \boldsymbol{\mathcal{B}}_{ss}=\boldsymbol{\mathcal{B}}_{ii}=\operatorname{diag}\{T,1,1,\ldots\},\nonumber\\
    \boldsymbol{\mathcal{B}}_{si}=\operatorname{diag}\{R,0,0,\ldots\},\;\boldsymbol{\mathcal{B}}_{is}=\operatorname{diag}\{-R,0,0,\ldots\},
\end{gather}
from the analysis of the Fabry-P\'{e}rot (FP) cavity, the new matrix for $\hat{\mathcal{U}}'$ is the Redheffer star product \cite{sapm1960391269,Li:96} of the two matrices $\boldsymbol{\mathcal{U}}'=\boldsymbol{\mathcal{U}}\star\boldsymbol{\mathcal{B}}$ (also unitary) whose submatrices are
\begin{align}\label{redhefferexp}
    &\boldsymbol{\mathcal{U}}'_{ss}=\boldsymbol{\mathcal{B}}_{ss}(\boldsymbol{I}-\boldsymbol{\mathcal{U}}_{si}\boldsymbol{\mathcal{B}}_{is})^{-1}\boldsymbol{\mathcal{U}}_{ss},\nonumber\\
    &\boldsymbol{\mathcal{U}}'_{si}=\boldsymbol{\mathcal{B}}_{si}+\boldsymbol{\mathcal{B}}_{ss}(\boldsymbol{I}-\boldsymbol{\mathcal{U}}_{si}\boldsymbol{\mathcal{B}}_{is})^{-1}\boldsymbol{\mathcal{U}}_{si}\boldsymbol{\mathcal{B}}_{ii},\nonumber\\
    &\boldsymbol{\mathcal{U}}'_{is}=\boldsymbol{\mathcal{U}}_{is}+\boldsymbol{\mathcal{U}}_{ii}(\boldsymbol{I}-\boldsymbol{\mathcal{B}}_{is}\boldsymbol{\mathcal{U}}_{si})^{-1}\boldsymbol{\mathcal{B}}_{is}\boldsymbol{\mathcal{U}}_{ss},\nonumber\\
    &\boldsymbol{\mathcal{U}}'_{ii}=\boldsymbol{\mathcal{U}}_{ii}(\boldsymbol{I}-\boldsymbol{\mathcal{B}}_{is}\boldsymbol{\mathcal{U}}_{si})^{-1}\boldsymbol{\mathcal{B}}_{ii},
\end{align}
where $\boldsymbol{I}$ is the identity matrix whose width is the \emph{s} or \emph{i} path number. Then, we use the mathematical induction again and prove the base case and the induction step (see Appendixes \ref{proof1} and \ref{proof2} for the proofs):

\textit{Lemma 1}. If $n_{i1}=0$, then
\begin{equation}\label{result}
    (\prescript{\mathbf{m}_s}{\mathbf{m}_i}{}|\hat{U}'|^{\mathbf{n}_s}_{\mathbf{n}_i})=\frac{C_UT}{1+R\mathcal{U}_{s1,i1}}\{\prescript{\mathbf{m}_s}{\mathbf{m}_i}{}|\hat{\mathcal{U}}'|^{\mathbf{n}_s}_{\mathbf{n}_i}\}.
\end{equation}

\textit{Lemma 2}. If Eq.\ \eqref{result} is true for a certain $n_{i1}$ (other quantities can be arbitrary as long as $\Delta\geq0$), then it is true for $n_{i1}\to n_{i1}+1$.

So, Eq.\ \eqref{result} is true for any $n_{i1}\geq0$, which means the new system $\hat{U}'$ satisfies the duality with the new NC $C_{U'}=C_UT/(1+R\mathcal{U}_{s1,i1})$. Then, our final duality theorem is: When considering only monochromatic \emph{s} and \emph{i} lights, an arbitrary quantum nonlinear optical system consisting of nondegenerate PDCs and linear systems satisfies the nonlinear-linear duality, and its PTR setup is a replacement of all the PDCs by hypothetical BSs with $R=\tanh r$. The NC is discussed as follows.

Given the explicit form $\hat{U}=\hat{U}_N\cdots\hat{U}_2\hat{U}_1$ of a nonlinear system, where $\hat{U}_j=\hat{L}_{sj}\hat{L}_{ij}\hat{G}_j$ and the corresponding $\hat{\mathcal{B}}_j$ of $\hat{G}_j$ has the amplitude transmittance $T_j$ and reflectance $R_j$ ($R_1=0$ and $T_1=1$ if the system begins with linear systems), as shown in Fig.\ \ref{ptrfig}(c), denoting
\begin{equation}
    \boldsymbol{\mathcal{U}}_j=\boldsymbol{\mathcal{B}}_j\star\begin{pmatrix}\boldsymbol{L}_{sj}&\\&\boldsymbol{L}_{ij}^\mathrm{T}\end{pmatrix}
    =\begin{pmatrix}\boldsymbol{L}_{sj}&\\&\boldsymbol{I}\end{pmatrix}\boldsymbol{\mathcal{B}}_j\begin{pmatrix}\boldsymbol{I}&\\&\boldsymbol{L}_{ij}^\mathrm{T}\end{pmatrix}
\end{equation}
and the intermediate results $\boldsymbol{\mathcal{U}}_{\star n}=\boldsymbol{\mathcal{U}}_1\star\boldsymbol{\mathcal{U}}_2\star\cdots\star\boldsymbol{\mathcal{U}}_n$ ($1\leq n\leq N$), the scattering matrix of the PTR setup $\boldsymbol{\mathcal{U}}=\boldsymbol{\mathcal{U}}_{\star N}$ and the NC $C_U=\beta T_1T_2\cdots T_n$, where
\begin{equation}\label{betacal}
    \beta=\prod_{n=1}^{N-1}\frac{1}{1+R_{n+1}(\boldsymbol{\mathcal{U}}_{\star n})_{s1,i1}}.
\end{equation}
Note that $(\boldsymbol{\mathcal{U}}_j)_{ss}=\boldsymbol{L}_{sj}(\boldsymbol{\mathcal{B}}_j)_{ss}$,
\begin{equation}
    \frac{1}{1+R_{n+1}(\boldsymbol{\mathcal{U}}_{\star n})_{s1,i1}}
    =\operatorname{det}[\boldsymbol{I}-(\boldsymbol{\mathcal{U}}_{\star n})_{si}(\boldsymbol{\mathcal{B}}_{n+1})_{is}]^{-1},
\end{equation}
\begin{align}
    \boldsymbol{\mathcal{U}}_{ss}
    &=\boldsymbol{L}_{sN}(\boldsymbol{\mathcal{B}}_N)_{ss}[\boldsymbol{I}-(\boldsymbol{\mathcal{U}}_{\star N-1})_{si}(\boldsymbol{\mathcal{B}}_N)_{is}]^{-1}\boldsymbol{L}_{s\,N-1}\cdots\nonumber\\
    &\qquad[\boldsymbol{I}-(\boldsymbol{\mathcal{U}}_{\star2})_{si}(\boldsymbol{\mathcal{B}}_3)_{is}]^{-1}\boldsymbol{L}_{s2}(\boldsymbol{\mathcal{B}}_2)_{ss}\nonumber\\
    &\qquad[\boldsymbol{I}-(\boldsymbol{\mathcal{U}}_{\star1})_{si}(\boldsymbol{\mathcal{B}}_2)_{is}]^{-1}\boldsymbol{L}_{s1}(\boldsymbol{\mathcal{B}}_1)_{ss},
\end{align}
$\operatorname{det}(\boldsymbol{\mathcal{B}}_j)_{ss}=T_j$, and $|\operatorname{det}\boldsymbol{L}_{sj}|=1$, so $|C_U|=|\operatorname{det}\boldsymbol{\mathcal{U}}_{ss}|$ and the argument difference is $\sum_{j=1}^N\arg\operatorname{det}\boldsymbol{L}_{sj}$ (a similar result with respect to $\boldsymbol{\mathcal{U}}_{ii}$ can be obtained). If $|C_U|=1$, $\boldsymbol{\mathcal{U}}_{ss}$ and $\boldsymbol{\mathcal{U}}_{ii}$ are unitary \cite{svd}, the PTR setup transmits all the lights, and the PDC processes are canceled.

If the initial and postselected states are other pure states, the two \emph{i} states in the PTR expression should be the complex conjugate of the original ones \cite{pnas.2010827117}
\begin{equation}\label{purestateposts}
    \langle\psi_s;\psi_i|\hat{U}|\psi_{s0};\psi_{i0}\rangle=C_U\langle\psi_s;\psi_{i0}^\ast|\hat{\mathcal{U}}|\psi_{s0};\psi_i^\ast\rangle.
\end{equation}
See Appendix \ref{geneproj} for the case of general state detection methods involving observable operators and a discussion on the presence of linear losses in the nonlinear system.

Till now, this theory concerns only monochromatic down-converted lights and optical systems insensitive to the polarization. Appendix \ref{polychrom} discusses this case by treating different frequencies as groups of spatial paths, and finds the \emph{i} light should travel backward in time in the PTR setup, which is exactly the meaning of PTR and reminiscent of Klyshko's original thought on the AWP \cite{KLYSHKO1988133,PhysRevA.110.063710}. If the polarization is considered, one spatial path becomes two to describe two orthogonal polarizations which can be linearly coupled by wave plates, and a nondegenerate type-II PDC acts on an \emph{s} and an \emph{i} path with orthogonal polarizations. However, if degenerate type-II PDC is used and PTR is performed with respect to the polarization, a wave plate, exactly the ``hypothetical'' BS, becomes a PDC, which is beyond our concern.

\section{Analyses and examples}\label{aae}

From Eqs.\ \eqref{insertbasisptr} and \eqref{result}, the connection between the postselection amplitude of the whole cavity system and the amplitude product of two systems summed over all possible intracavity states
\begin{align}
    &\sum_{n'=0}^{+\infty}\{\prescript{\mathbf{m}_s}{\mathbf{m}_i}{}|\hat{\mathcal{U}}|^{\Delta+n',\mathbf{n}_{s+}}_{n',\mathbf{n}_{i+}}\}\{\prescript{\Delta+n'}{n'}{}|\hat{\mathcal{B}}|^{n_{s1}}_{n_{i1}}\}\nonumber\\
    =\,&\frac{\{\prescript{\mathbf{m}_s}{\mathbf{m}_i}{}|\hat{\mathcal{U}}'|^{\mathbf{n}_s}_{\mathbf{n}_i}\}}{1+R\mathcal{U}_{s1,i1}}=\{\prescript{\mathbf{m}_s}{\mathbf{m}_i}{}|\hat{\mathcal{U}}'|^{\mathbf{n}_s}_{\mathbf{n}_i}\}\sum_{n=0}^{+\infty}(-R\mathcal{U}_{s1,i1})^n
\end{align}
has a possible intuitive explanation. $(-R\mathcal{U}_{s1,i1})^n$ is the amplitude of $n$ photons traveling a loop inside the cavity, so the first line of this equation contains not only the true contribution from $\{\prescript{\mathbf{m}_s}{\mathbf{m}_i}{}|\hat{\mathcal{U}}'|^{\mathbf{n}_s}_{\mathbf{n}_i}\}$ where there is initially no photon inside the cavity, but also the cases where there is one or more photons inside it which remain inside after postselection. 

\begin{figure}[b]
\includegraphics[width=0.48\textwidth]{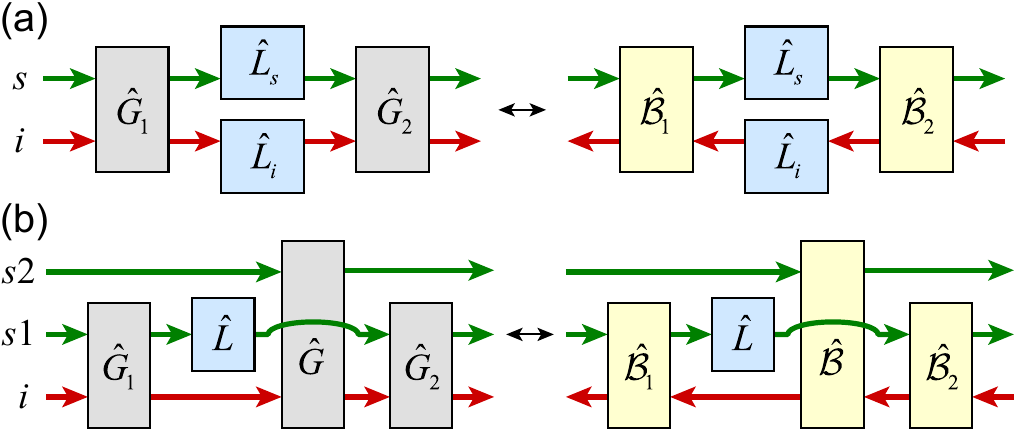}
\caption{\label{ptrexfig}Examples of the duality. (a) Two cascaded PDCs with two phase plates (the only possible type of single-path linear lossless setups) inserted. (b) The PDC $\hat{G}_1$ acting on paths $s1$ and $i$, a $\pi$ phase shift on $s1$, $\hat{G}$ acting on $s2$ and $i$, and $\hat{G}_2$ acting on $s1$ and $i$.}
\end{figure}
We present a simple single-path example, two cascaded PDCs with $r_1$ and $r_2$. The PTR setup is a cavity made of two hypothetical BSs with $R_1=\tanh r_1$ and $R_2=\tanh r_2$. With no phase between them, the overall amplitude reflectance is $(R_1+R_2)/(1+R_1R_2)=\tanh(r_1+r_2)$ ($i\to s$; adding a minus sign for $s\to i$) from the FP cavity calculation and the NC is $T_1T_2/(1+R_1R_2)=\operatorname{sech}(r_1+r_2)$, which indeed describes the PTR setup of a PDC with $r_1+r_2$. Then, if $r_1=r_2=r$, inserting two phase plates $\Phi_s$ and $\Phi_i$ at the \emph{s} and \emph{i} path respectively between the PDCs [as shown in Fig.\ \ref{ptrexfig}(a), becoming an SU(1,1) interferometer \cite{PhysRevA.33.4033,Chekhova:16,5.0004873}], letting $\Phi=\Phi_s+\Phi_i$, $\beta=1/(1+R^2e^{i\Phi})$, $C_U=\beta T^2$, and the explicit form of $\boldsymbol{\mathcal{U}}$,
\begin{equation}\label{su11ptr}
    \begin{pmatrix}\beta T^2e^{i\Phi_s}&2\beta R\cos(\Phi/2)e^{i\Phi/2}\\
    -2\beta R\cos(\Phi/2)e^{i\Phi/2}&\beta T^2e^{i\Phi_i}\end{pmatrix},
\end{equation}
can also be constructed by a hypothetical BS with the amplitude transmittance $|\beta|T^2$ (which is also the NC of this setup) surrounded by four phase plates \cite{fourphase}. So, although different linear setups may have the same $\boldsymbol{\mathcal{U}}$ (and $|C_U|$), the presence of cavities brings a phase to $\beta$ and the NC. When $\Phi=\pi$, the two PDCs are canceled, the cavity transmits all the lights, and $C_U=1$. If $r_1\neq r_2$, the cancellation is impossible.

Another example with two \emph{s} paths is shown in Fig.\ \ref{ptrexfig}(b). After the first PDC $\hat{G}_1$ acting on $s1$ and $i$, a $\pi$ phase shift is on $s1$ and a PDC $\hat{G}$ acts on $s2$ and $i$. Finally, $\hat{G}_2$ acts on $s1$ and $i$. If $r_1>r_2$ and $T=R_2/R_1$ (i.e., $\cosh r\tanh r_2=\tanh r_1$), we have $C_U=T_1T/T_2$ and
\begin{equation}
    \boldsymbol{\mathcal{U}}=\frac{1}{T_2^2}\left(\begin{array}{cc|c}
        -T_1T_2&RR_1T_2&0\\
        RR_2T_1&TT_1^2&RT_2\\
        \hline
        -R^2R_1&-RT_1&TT_1T_2
    \end{array}\right)
\end{equation}
(the order is $s1,s2,i$, e.g., $\mathcal{U}_{s1,i}=0$), which means for the vacuum \emph{s} input, the final state of path $s1$ is still vacuum, and the down-converted photons are purely on $s2$ and $i$. So, when $r_1>r_2$, the light on $s1$ can still be canceled by inserting a PDC not acting on this path. This can be verified by checking the Bogoliubov transformation of $\hat{a}_{s1}$ does not have the $\hat{a}_i^\dagger$ term and $\langle\mathrm{vac};\psi_{i0}|\hat{U}^\dagger\hat{a}_{s1}^\dagger\hat{a}_{s1}\hat{U}|\mathrm{vac};\psi_{i0}\rangle=0$.

If the \emph{s} input and both the postselections of a nonlinear system are multimode coherent states $\boldsymbol{\alpha}_s$, $\boldsymbol{\beta}_s$, and $\boldsymbol{\beta}_i$, the outputs of its PTR setup from $\boldsymbol{\alpha}_s$ and $\boldsymbol{\beta}_i^\ast$ are still coherent which can be calculated classically. Its inner product with $\boldsymbol{\beta}_s$ and the conjugated \emph{i} initial state is easy to calculate. As the postselection probability of the nonlinear system is proportional to its \emph{Q}-function at $Q(\boldsymbol{\beta}_s,\boldsymbol{\beta}_i)$, this duality provides a way to calculate it when one of the two input states is coherent. If the \emph{i} input is also coherent $\boldsymbol{\alpha}_i$, the initial state is Gaussian, the nonlinear system performs a Gaussian operation (without displacement) which can be described by a symplectic matrix \cite{Leonhardt_2003,RevModPhys.84.621,RevModPhys.92.035005,PhysRevA.98.062314,Quesada:22}, and $Q(\boldsymbol{\beta}_s,\boldsymbol{\beta}_i)$ of the final state can be calculated from the Gaussian state parameters \cite{PhysRevA.100.032326,PhysRevA.99.053816}. Meanwhile, because of the duality, the \emph{Q}-function of the final state of the PTR setup $Q_\mathrm{PTR}(\boldsymbol{\beta}_s,\boldsymbol{\alpha}_i^\ast)$ should be $Q(\boldsymbol{\beta}_s,\boldsymbol{\beta}_i)/|C_U|^2$. In Appendix \ref{gaussduality}, we introduce the transfer matrix of the PTR setup and use the Gaussian state and operation approach to prove this \emph{Q}-function relation. Then, using the Glauber-Sudarshan \emph{P}-function, the duality can also be proved (except $\arg C_U$ which is usually meaningless). Finally, we discuss why we cannot easily derive a duality if single-mode squeezing (often realized by degenerate type-I PDC) is present.

\section{Discussion}

In this article, based on the multimode Fock state, we proved the nonlinear-linear duality for quantum nonlinear optical systems consisting of nondegenerate PDCs and linear systems. The PTR setup which often involves cavities is from directly replacing PDCs in the original nonlinear setup with hypothetical BSs. So, rather than decomposing a nonlinear process into an equivalent setup of a specific form \cite{PhysRevA.98.062314}, this duality preserves the geometry of the original setup. An alternative proof using the Gaussian state theory is presented in Appendix \ref{gaussduality}. A shortcoming is this duality cannot take single-mode squeezing or real optical cavities \cite{realcavity} into account. Also, the calculation of the postselection amplitude after a linear system from a multiphoton initial state is known to be hard (boson sampling \cite{science.1231692,Tillmann2013}), and so is a nonlinear one. Nevertheless, one possible application about the \emph{Q}-function calculation has been discussed.

This theory is based on discrete spatial paths, which are usually seen in integrated photonic devices \cite{Wang2020,Quesada:22}, but we believe it can be applied to the continuous-variable spatial degree of freedom and, thus, quantum imaging by modifying the expressions from path indices to positions. The light propagation in media and the phase of the pump beam need to be considered as infinitesimal linear systems, so the PTR setup should be solved from coupled wave equations. If the low-gain approximation is used for nondegenerate SPDC ($\mathbf{m}_s=\mathbf{m}_i=\mathbf{0}$) and $\|\mathbf{n}_s\|=\|\mathbf{n}_i\|=1$, the light fields reflected more than once by hypothetical BSs of low reflectance are ignored (the cavity issue no longer exists), and our theorem becomes the AWP \cite{PhysRevA.110.063710}. In fact, to ensure two-fold coincidence counting roughly measures the two-photon detection probability (without higher-order SPDC terms), from our analysis, we only need $|C_U|\approx1$, so the gain values of all the PDCs need not be low. Explicitly, $(\prescript{\mathbf{0}}{\mathbf{0}}{}|\hat{U}|^{\mathbf{e}_{sk}}_{\mathbf{e}_{ij}})\approx\{\prescript{\mathbf{0}}{\mathbf{0}}{}|\hat{\mathcal{U}}|^{\mathbf{e}_{sk}}_{\mathbf{e}_{ij}}\}=\mathcal{U}_{sk,ij}$ ($\mathbf{e}_{sk}$ and $\mathbf{e}_{ij}$ are unit vectors), i.e., the biphoton postselection amplitude equals a scattering matrix element. An interesting result is, if $\|\mathbf{n}_s\|=\|\mathbf{n}_i\|=2$, the four-photon postselection amplitude from SPDC also has an AWP theory where the setup is purely classical (see Appendix \ref{fourawp} for its description and an example, the quantum teleportation experiment \cite{Bouwmeester1997}).

\begin{acknowledgments}
We thank Xiang-Fa Zhou for the useful discussion. This work was supported by the Innovation Program for Quantum Science and Technology (Grants No.\ 2021ZD0301200 and No.\ 2021ZD0301400), National Natural Science Foundation of China (Grants No.\ 92365205, No.\ W2411001), and USTC Major Frontier Research Program (No.\ LS2030000002).

\section*{Data availability}
No data were created or analyzed in this study.
\end{acknowledgments}

\appendix

\section{Proof of Lemma 1}\label{proof1}

When $n_{i1}=0$, an input port of $\hat{\mathcal{B}}$ is empty and we have
\begin{equation}
    \{\prescript{\Delta+n'}{n'}{}|\hat{\mathcal{B}}|^{\Delta}_{0}\}=\sqrt{\frac{(\Delta+n')!}{\Delta!n'!}}T^{\Delta}(-R)^{n'}
\end{equation}
according to the binomial distribution. For $\hat{\mathcal{U}}$, we have
\begin{widetext}
\begin{align}
    \{\prescript{\mathbf{m}_s}{\mathbf{m}_i}{}|\hat{\mathcal{U}}|^{\Delta+n',\mathbf{n}_{s+}}_{n',\mathbf{n}_{i+}}\}
    &=\langle\Delta+n',\mathbf{n}_{s+};\mathbf{m}_i|\hat{\mathcal{U}}|\mathbf{m}_s;n',\mathbf{n}_{i+}\rangle
    =\sqrt{\frac{\Delta!}{n'!(\Delta+n')!}}\langle\Delta,\mathbf{n}_{s+};\mathbf{m}_i|\hat{a}_{s1}^{n'}\hat{\mathcal{U}}(\hat{a}_{i1}^\dagger)^{n'}|\mathbf{m}_s;0,\mathbf{n}_{i+}\rangle\nonumber\\
    &=\sqrt{\frac{\Delta!}{n'!(\Delta+n')!}}\langle\Delta,\mathbf{n}_{s+};\mathbf{m}_i|\hat{\mathcal{U}}\big(\hat{\mathcal{U}}^\dagger\hat{a}_{s1}\hat{\mathcal{U}}\big)^{n'}(\hat{a}_{i1}^\dagger)^{n'}|\mathbf{m}_s;0,\mathbf{n}_{i+}\rangle\nonumber\\
    &=\sqrt{\frac{\Delta!}{n'!(\Delta+n')!}}\langle\Delta,\mathbf{n}_{s+};\mathbf{m}_i|\hat{\mathcal{U}}\Bigg[\sum_j(\mathcal{U}_{s1,sj}\hat{a}_{sj}+\mathcal{U}_{s1,ij}\hat{a}_{ij})\Bigg]^{n'}(\hat{a}_{i1}^\dagger)^{n'}|\mathbf{m}_s;0,\mathbf{n}_{i+}\rangle,
\end{align}
so,
\begin{equation}\label{theo2}
    \sum_{n'=0}^{+\infty}\{\prescript{\mathbf{m}_s}{\mathbf{m}_i}{}|\hat{\mathcal{U}}|^{\Delta+n',\mathbf{n}_{s+}}_{n',\mathbf{n}_{i+}}\}\{\prescript{\Delta+n'}{n'}{}|\hat{\mathcal{B}}|^{\Delta}_{0}\}
    =\sum_{n'=0}^{+\infty}\frac{T^{\Delta}}{n'!}\langle\Delta,\mathbf{n}_{s+};\mathbf{m}_i|\hat{\mathcal{U}}\Bigg[-R\sum_j(\mathcal{U}_{s1,sj}\hat{a}_{sj}+\mathcal{U}_{s1,ij}\hat{a}_{ij})\Bigg]^{n'}(\hat{a}_{i1}^\dagger)^{n'}|\mathbf{m}_s;0,\mathbf{n}_{i+}\rangle.
\end{equation}

From Eq.\ \eqref{insertbasisptr}, we need to prove
\begin{equation}\label{s7}
    \frac{1}{1+R\mathcal{U}_{s1,i1}}\{\prescript{\mathbf{m}_s}{\mathbf{m}_i}{}|\hat{\mathcal{U}}'|^{\mathbf{n}_s}_{\mathbf{n}_i}\}
    \stackrel{?}{=}\sum_{n'=0}^{+\infty}\frac{T^{\Delta}}{n'!}\langle\Delta,\mathbf{n}_{s+};\mathbf{m}_i|\hat{\mathcal{U}}\Bigg[-R\sum_j(\mathcal{U}_{s1,sj}\hat{a}_{sj}+\mathcal{U}_{s1,ij}\hat{a}_{ij})\Bigg]^{n'}(\hat{a}_{i1}^\dagger)^{n'}|\mathbf{m}_i;0,\mathbf{n}_{i+}\rangle.
\end{equation}
First, we have
\begin{equation}
    \hat{\mathcal{U}}'\hat{a}_{ij}^\dagger\hat{\mathcal{U}}'^\dagger
    =\sum_k\big(\mathcal{U}'_{sk,ij}\hat{a}_{sk}^\dagger+\mathcal{U}'_{ik,ij}\hat{a}_{ik}^\dagger\big).
\end{equation}
From Eq.\ \eqref{redhefferexp} and the fact that $\boldsymbol{\mathcal{B}}_{is}=-R\mathbf{e}_{i1}\mathbf{e}_{s1}^\mathrm{T}$ ($\mathbf{e}_{i1}$ and $\mathbf{e}_{s1}$ are unit column vectors), we have, for example, when $j>1$, 
\begin{align}
    \boldsymbol{\mathcal{U}}'_{si}\mathbf{e}_{ij}
    &=\sum_{n=0}^{+\infty}\boldsymbol{\mathcal{B}}_{ss}(\boldsymbol{\mathcal{U}}_{si}\boldsymbol{\mathcal{B}}_{is})^n\boldsymbol{\mathcal{U}}_{si}\mathbf{e}_{ij}
    =\boldsymbol{\mathcal{B}}_{ss}\boldsymbol{\mathcal{U}}_{si}\mathbf{e}_{ij}-R\sum_{n=0}^{+\infty}(-R)^n\boldsymbol{\mathcal{B}}_{ss}\boldsymbol{\mathcal{U}}_{si}\mathbf{e}_{i1}\big(\mathbf{e}_{s1}^\mathrm{T}\boldsymbol{\mathcal{U}}_{si}\mathbf{e}_{i1}\big)^n\mathbf{e}_{s1}^\mathrm{T}\boldsymbol{\mathcal{U}}_{si}\mathbf{e}_{ij}\nonumber\\
    &=\boldsymbol{\mathcal{B}}_{ss}\left[\boldsymbol{\mathcal{U}}_{si}\mathbf{e}_{ij}-R\mathcal{U}_{s1,ij}\sum_{n=0}^{+\infty}(-R\mathcal{U}_{s1,i1})^n\boldsymbol{\mathcal{U}}_{si}\mathbf{e}_{i1}\right]
    =\boldsymbol{\mathcal{B}}_{ss}\left(\boldsymbol{\mathcal{U}}_{si}\mathbf{e}_{ij}-\frac{R\mathcal{U}_{s1,ij}}{1+R\mathcal{U}_{s1,i1}}\boldsymbol{\mathcal{U}}_{si}\mathbf{e}_{i1}\right),
\end{align}
and the matrix elements
\begin{gather}
    \mathcal{U}'_{s1,ij}=\mathbf{e}_{s1}^\mathrm{T}\boldsymbol{\mathcal{U}}'_{si}\mathbf{e}_{ij}=T\left(\mathcal{U}_{s1,ij}-\frac{R\mathcal{U}_{s1,ij}}{1+R\mathcal{U}_{s1,i1}}\mathcal{U}_{s1,i1}\right)\;\textrm{($j>1$, not fully simplified on purpose)},\nonumber\\
    \mathcal{U}'_{sk,ij}=\mathcal{U}_{sk,ij}-\frac{R\mathcal{U}_{s1,ij}}{1+R\mathcal{U}_{s1,i1}}\mathcal{U}_{sk,i1}\;\textrm{($j>1$, $k>1$)},\;
    \mathcal{U}'_{ik,ij}=\mathcal{U}_{ik,ij}-\frac{R\mathcal{U}_{s1,ij}}{1+R\mathcal{U}_{s1,i1}}\mathcal{U}_{ik,i1}\;\textrm{($j>1$)},\nonumber\\
    \mathcal{U}'_{s1,sj}=T\left(\mathcal{U}_{s1,sj}-\frac{R\mathcal{U}_{s1,sj}}{1+R\mathcal{U}_{s1,i1}}\mathcal{U}_{s1,i1}\right)\;\textrm{(not fully simplified on purpose)},\nonumber\\
    \mathcal{U}'_{sk,sj}=\mathcal{U}_{sk,sj}-\frac{R\mathcal{U}_{s1,sj}}{1+R\mathcal{U}_{s1,i1}}\mathcal{U}_{sk,i1}\;\textrm{($k>1$)},\;
    \mathcal{U}'_{ik,sj}=\mathcal{U}_{ik,sj}-\frac{R\mathcal{U}_{s1,sj}}{1+R\mathcal{U}_{s1,i1}}\mathcal{U}_{ik,i1}.
\end{gather}
Letting $\tilde{\mathcal{U}}'_{s1,ij}=\mathcal{U}'_{s1,ij}/T$ ($j>1$), $\tilde{\mathcal{U}}'_{sk,ij}=\mathcal{U}'_{sk,ij}$ ($j>1$ and $k>1$), $\tilde{\mathcal{U}}'_{s1,sj}=\mathcal{U}'_{s1,sj}/T$, and $\tilde{\mathcal{U}}'_{sk,sj}=\mathcal{U}'_{sk,sj}$ ($k>1$), we have
\begin{align}
    \hat{\mathcal{U}}\left(\hat{a}_{ij}^\dagger-\frac{R\mathcal{U}_{s1,ij}}{1+R\mathcal{U}_{s1,i1}}\hat{a}_{i1}^\dagger\right)\hat{\mathcal{U}}^\dagger
    &=\sum_k\left[\left(\mathcal{U}_{sk,ij}-\frac{R\mathcal{U}_{s1,ij}}{1+R\mathcal{U}_{s1,i1}}\mathcal{U}_{sk,i1}\right)\hat{a}_{sk}^\dagger+\left(\mathcal{U}_{ik,ij}-\frac{R\mathcal{U}_{s1,ij}}{1+R\mathcal{U}_{s1,i1}}\mathcal{U}_{ik,i1}\right)\hat{a}_{ik}\right]\nonumber\\
    &=\sum_k\big(\tilde{\mathcal{U}}'_{sk,ij}\hat{a}_{sk}^\dagger+\mathcal{U}'_{ik,ij}\hat{a}_{ik}^\dagger\big)
\end{align}
for $j>1$ (the sum is over all $k\ge1$) and
\begin{equation}
    \hat{\mathcal{U}}\left(\hat{a}_{sj}^\dagger-\frac{R\mathcal{U}_{s1,sj}}{1+R\mathcal{U}_{s1,i1}}\hat{a}_{i1}^\dagger\right)\hat{\mathcal{U}}^\dagger
    =\sum_k\big(\tilde{\mathcal{U}}'_{sk,sj}\hat{a}_{sk}^\dagger+\mathcal{U}'_{ik,sj}\hat{a}_{ik}^\dagger\big)
\end{equation}
for $j\geq1$. Then,
\begin{align}
    &\{\prescript{\mathbf{m}_s}{\mathbf{m}_i}{}|\hat{\mathcal{U}}'|^{\mathbf{n}_s}_{\mathbf{n}_i}\}\nonumber\\
    =\,&\langle\Delta,\mathbf{n}_{s+};\mathbf{m}_i|\hat{\mathcal{U}}'|\mathbf{m}_s;0,\mathbf{n}_{i+}\rangle
    =\langle\Delta,\mathbf{n}_{s+};\mathbf{m}_i|\hat{\mathcal{U}}'\left[\prod_{j}\frac{(a_{sj}^\dagger)^{m_{sj}}}{\sqrt{m_{sj}!}}\right]\left[\prod_{j>1}\frac{(a_{ij}^\dagger)^{n_{ij}}}{\sqrt{n_{ij}!}}\right]|\mathrm{vac}\rangle\nonumber\\
    =\,&\langle\Delta,\mathbf{n}_{s+};\mathbf{m}_i|\left\{\prod_{j}\frac{1}{\sqrt{m_{sj}!}}\left[\sum_k\big(\mathcal{U}'_{sk,sj}\hat{a}_{sk}^\dagger+\mathcal{U}'_{ik,sj}\hat{a}_{ik}^\dagger\big)\right]^{m_{sj}}\right\}\left\{\prod_{j>1}\frac{1}{\sqrt{n_{ij}!}}\left[\sum_k\big(\mathcal{U}'_{sk,ij}\hat{a}_{sk}^\dagger+\mathcal{U}'_{ik,ij}\hat{a}_{ik}^\dagger\big)\right]^{n_{ij}}\right\}|\mathrm{vac}\rangle\nonumber\\
    =\,&T^{\Delta}\langle\Delta,\mathbf{n}_{s+};\mathbf{m}_i|\left\{\prod_{j}\frac{1}{\sqrt{m_{sj}!}}\left[\sum_k\big(\tilde{\mathcal{U}}'_{sk,sj}\hat{a}_{sk}^\dagger+\mathcal{U}'_{ik,sj}\hat{a}_{ik}^\dagger\big)\right]^{m_{sj}}\right\}\left\{\prod_{j>1}\frac{1}{\sqrt{n_{ij}!}}\left[\sum_k\big(\tilde{\mathcal{U}}'_{sk,ij}\hat{a}_{sk}^\dagger+\mathcal{U}'_{ik,ij}\hat{a}_{ik}^\dagger\big)\right]^{n_{ij}}\right\}|\mathrm{vac}\rangle\nonumber\\
    =\,&T^{\Delta}\langle\Delta,\mathbf{n}_{s+};\mathbf{m}_i|\hat{\mathcal{U}}\overbrace{\left[\prod_{j}\frac{1}{\sqrt{m_{sj}!}}\left(\hat{a}_{sj}^\dagger-\frac{R\mathcal{U}_{s1,sj}}{1+R\mathcal{U}_{s1,i1}}\hat{a}_{i1}^\dagger\right)^{m_{sj}}\right]\left[\prod_{j>1}\frac{1}{\sqrt{n_{ij}!}}\left(\hat{a}_{ij}^\dagger-\frac{R\mathcal{U}_{s1,ij}}{1+R\mathcal{U}_{s1,i1}}\hat{a}_{i1}^\dagger\right)^{n_{ij}}\right]|\mathrm{vac}\rangle}^\textrm{Let it be $|\tilde{\psi}(\mathbf{m}_s;\mathbf{n}_{i+})\rangle$ (unnormalized)},
\end{align}
where the fourth line is obtained because path $s1$ is postselected to $\Delta$ photons.

Finally, we need to prove
\begin{equation}\label{provbyind}
    \frac{1}{1+R\mathcal{U}_{s1,i1}}|\tilde{\psi}(\mathbf{m}_s;\mathbf{n}_{i+})\rangle
    \stackrel{?}{=}\sum_{n'=0}^{+\infty}\frac{1}{n'!}\Bigg[-R\sum_j(\mathcal{U}_{s1,sj}\hat{a}_{sj}+\mathcal{U}_{s1,ij}\hat{a}_{ij})\Bigg]^{n'}(\hat{a}_{i1}^\dagger)^{n'}|\mathbf{m}_s;0,\mathbf{n}_{i+}\rangle,
\end{equation}
and we do this by mathematical induction (for the third time). For simplicity, we use $\hat{\mathcal{U}}^\dagger\hat{a}_{s1}\hat{\mathcal{U}}$ to represent $\sum_j(\mathcal{U}_{s1,sj}\hat{a}_{sj}+\mathcal{U}_{s1,ij}\hat{a}_{ij})$. First, Eq.\ \eqref{provbyind} satisfies when $\mathbf{m}_s=\mathbf{n}_i=\mathbf{0}$ because
\begin{equation}
    \sum_{n'=0}^{+\infty}\frac{1}{n'!}(-R\hat{\mathcal{U}}^\dagger\hat{a}_{s1}\hat{\mathcal{U}})^{n'}(\hat{a}_{i1}^\dagger)^{n'}|\mathrm{vac}\rangle
    =\sum_{n=0}^{+\infty}\frac{1}{n!}(-R\mathcal{U}_{s1,i1})^n\hat{a}_{i1}^n(\hat{a}_{i1}^\dagger)^n|\mathrm{vac}\rangle
    =\frac{1}{1+R\mathcal{U}_{s1,i1}}|\mathrm{vac}\rangle.
\end{equation}
If it satisfies for a certain $\mathbf{m}_s$ and $\mathbf{n}_{i+}$, for $\mathbf{m}_s$ and $\mathbf{n}_{i+}+\mathbf{e}_{ip}$ ($p>1$),
\begin{align}\label{s14}
    &\sum_{n'=0}^{+\infty}\frac{1}{n'!}(-R\hat{\mathcal{U}}^\dagger\hat{a}_{s1}\hat{\mathcal{U}})^{n'}(\hat{a}_{i1}^\dagger)^{n'}|\mathbf{m}_s;0,\mathbf{n}_{i+}+\mathbf{e}_{ip}\rangle\nonumber\\
    =\,&\frac{1}{\sqrt{n_{ip}+1}}\sum_{n'=0}^{+\infty}\frac{1}{n'!}\overbrace{(-R\hat{\mathcal{U}}^\dagger\hat{a}_{s1}\hat{\mathcal{U}})^{n'}\hat{a}_{ip}^\dagger}^\textrm{Reference when iterating}(\hat{a}_{i1}^\dagger)^{n'}|\mathbf{m}_s;0,\mathbf{n}_{i+}\rangle\nonumber\\
    =\,&\frac{1}{\sqrt{n_{ip}+1}}\sum_{n'=0}^{+\infty}\frac{1}{n'!}\big[\hat{a}_{ip}^\dagger(-R\hat{\mathcal{U}}^\dagger\hat{a}_{s1}\hat{\mathcal{U}})^{n'}-n'R\mathcal{U}_{s1,ip}(-R\hat{\mathcal{U}}^\dagger\hat{a}_{s1}\hat{\mathcal{U}})^{n'-1}\big](\hat{a}_{i1}^\dagger)^{n'}|\mathbf{m}_s;0,\mathbf{n}_{i+}\rangle\nonumber\\
    =\,&\frac{1}{\sqrt{n_{ip}+1}}\sum_{n'=0}^{+\infty}\frac{1}{n'!}\big[\hat{a}_{ip}^\dagger(-R\hat{\mathcal{U}}^\dagger\hat{a}_{s1}\hat{\mathcal{U}})^{n'}-R\mathcal{U}_{s1,ip}\overbrace{(-R\hat{\mathcal{U}}^\dagger\hat{a}_{s1}\hat{\mathcal{U}})^{n'}\hat{a}_{i1}^\dagger}^\textrm{To be iterated}\big](\hat{a}_{i1}^\dagger)^{n'}|\mathbf{m}_s;0,\mathbf{n}_{i+}\rangle,
\end{align}
where the fact that $(-R\hat{\mathcal{U}}^\dagger\hat{a}_{s1}\hat{\mathcal{U}})\hat{a}_{ip}^\dagger=\hat{a}_{ip}^\dagger(-R\hat{\mathcal{U}}^\dagger\hat{a}_{s1}\hat{\mathcal{U}})-R\mathcal{U}_{s1,ip}$
(from the commutation relation; also true when $p=1$) is used $n'$ times and the substitution $n'-1\to n'$ is applied to obtain the second term in the brackets in the last line. After infinite iterations, the remaining term where $\hat{a}_{i1}^\dagger$ is on the right side of $(-R\hat{\mathcal{U}}^\dagger\hat{a}_{s1}\hat{\mathcal{U}})^{n'}$ is infinitely small, so Eq.\ \eqref{s14} becomes
\begin{align}
    &\frac{1}{\sqrt{n_{ip}+1}}\left[\hat{a}_{ip}^\dagger-R\mathcal{U}_{s1,ip}\sum_{n=0}^{+\infty}(-R\mathcal{U}_{s1,i1})^n\hat{a}_{i1}^\dagger\right]\sum_{n'=0}^{+\infty}\frac{1}{n'!}(-R\hat{\mathcal{U}}^\dagger\hat{a}_{s1}\hat{\mathcal{U}})^{n'}(\hat{a}_{i1}^\dagger)^{n'}|\mathbf{m}_s;0,\mathbf{n}_{i+}\rangle\nonumber\\
    =\,&\frac{1}{1+R\mathcal{U}_{s1,i1}}\frac{1}{\sqrt{n_{ip}+1}}\left(\hat{a}_{ip}^\dagger-\frac{R\mathcal{U}_{s1,ip}}{1+R\mathcal{U}_{s1,i1}}\hat{a}_{i1}^\dagger\right)|\tilde{\psi}(\mathbf{m}_s;\mathbf{n}_{i+})\rangle,
\end{align}
which is the first line of Eq.\ \eqref{provbyind} for $\mathbf{m}_s$ and $\mathbf{n}_{i+}+\mathbf{e}_{ip}$. For $\mathbf{m}_s+\mathbf{e}_{sp}$ and $\mathbf{n}_{i+}$ ($p\geq1$), similarly,
\begin{align}
    &\sum_{n'=0}^{+\infty}\frac{1}{n'!}(-R\hat{\mathcal{U}}^\dagger\hat{a}_{s1}\hat{\mathcal{U}})^{n'}(\hat{a}_{i1}^\dagger)^{n'}|\mathbf{m}_s+\mathbf{e}_{sp};0,\mathbf{n}_{i+}\rangle\nonumber\\
    =\,&\frac{1}{\sqrt{m_{sp}+1}}\sum_{n'=0}^{+\infty}\frac{1}{n'!}\big[\hat{a}_{sp}^\dagger(-R\hat{\mathcal{U}}^\dagger\hat{a}_{s1}\hat{\mathcal{U}})^{n'}-R\mathcal{U}_{s1,sp}(-R\hat{\mathcal{U}}^\dagger\hat{a}_{s1}\hat{\mathcal{U}})^{n'}\hat{a}_{i1}^\dagger\big](\hat{a}_{i1}^\dagger)^{n'}|\mathbf{m}_s;0,\mathbf{n}_{i+}\rangle\nonumber\\
    =\,&\frac{1}{1+R\mathcal{U}_{s1,i1}}\frac{1}{\sqrt{m_{sp}+1}}\left(\hat{a}_{sp}^\dagger-\frac{R\mathcal{U}_{s1,sp}}{1+R\mathcal{U}_{s1,i1}}\hat{a}_{i1}^\dagger\right)|\tilde{\psi}(\mathbf{m}_s;\mathbf{n}_{i+})\rangle,
\end{align}
which is the first line of Eq.\ \eqref{provbyind} for $\mathbf{m}_s+\mathbf{e}_{sp}$ and $\mathbf{n}_{i+}$. So, Lemma 1 is proved.

\section{Proof of Lemma 2}\label{proof2}

First, we have
\begin{align}
    \{\prescript{\mathbf{m}_s}{\mathbf{m}_i}{}|\hat{\mathcal{U}}'|^{n_{i1}+1+\Delta,\mathbf{n}_{s+}}_{n_{i1}+1,\mathbf{n}_{i+}}\}
    &=\frac{1}{\sqrt{n_{i1}+1}}\langle n_{i1}+1+\Delta,\mathbf{n}_{s+};\mathbf{m}_i|\big(\hat{\mathcal{U}}'\hat{a}_{i1}^\dagger\hat{\mathcal{U}}'^\dagger\big)\hat{\mathcal{U}}'|\mathbf{m}_s;n_{i1},\mathbf{n}_{i+}\rangle\nonumber\\
    &=\frac{1}{\sqrt{n_{i1}+1}}\langle n_{i1}+1+\Delta,\mathbf{n}_{s+};\mathbf{m}_i|\sum_k\big(\mathcal{U}'_{sk,i1}\hat{a}_{sk}^\dagger+\mathcal{U}'_{ik,i1}\hat{a}_{ik}^\dagger\big)\hat{\mathcal{U}}'|\mathbf{m}_s;n_{i1},\mathbf{n}_{i+}\rangle\nonumber\\
    &=\sqrt{\frac{n_{i1}+1+\Delta}{n_{i1}+1}}\mathcal{U}'_{s1,i1}\{\prescript{\mathbf{m}_s}{\mathbf{m}_i}{}|\hat{\mathcal{U}}'|^{n_{i1}+\Delta,\mathbf{n}_{s+}}_{n_{i1},\mathbf{n}_{i+}}\}+\sum_{k>1}\sqrt{\frac{n_{sk}}{n_{i1}+1}}\mathcal{U}'_{sk,i1}\{\prescript{\mathbf{m}_s}{\mathbf{m}_i}{}|\hat{\mathcal{U}}'|^{n_{i1}+1+\Delta,\mathbf{n}_{s+}-\mathbf{e}_{sk}}_{n_{i1},\mathbf{n}_{i+}}\}\nonumber\\
    &\qquad\qquad+\sum_k\sqrt{\frac{m_{ik}}{n_{i1}+1}}\mathcal{U}'_{ik,i1}\{\prescript{\mathbf{m}_s}{\mathbf{m}_i-\mathbf{e}_{ik}}{}|\hat{\mathcal{U}}'|^{n_{i1}+1+\Delta,\mathbf{n}_{s+}}_{n_{i1},\mathbf{n}_{i+}}\},
\end{align}
where the matrix elements
\begin{equation}
    \mathcal{U}'_{s1,i1}=R+\frac{T^2\mathcal{U}_{s1,i1}}{1+R\mathcal{U}_{s1,i1}},\;
    \mathcal{U}'_{sk,i1}=\frac{T\mathcal{U}_{sk,i1}}{1+R\mathcal{U}_{s1,i1}}\;\textrm{($k>1$)},\;
    \mathcal{U}'_{ik,i1}=\frac{T\mathcal{U}_{ik,i1}}{1+R\mathcal{U}_{s1,i1}}\;\textrm{($k\geq1$)}.
\end{equation}
Then,
\begin{align}\label{s21}
    &\frac{C_UT}{1+R\mathcal{U}_{s1,i1}}\{\prescript{\mathbf{m}_s}{\mathbf{m}_i}{}|\hat{\mathcal{U}}'|^{n_{i1}+1+\Delta,\mathbf{n}_{s+}}_{n_{i1}+1,\mathbf{n}_{i+}}\}\nonumber\\
    =\,&\sqrt{\frac{n_{i1}+1+\Delta}{n_{i1}+1}}\mathcal{U}'_{s1,i1}(\prescript{\mathbf{m}_s}{\mathbf{m}_i}{}|\hat{U}'|^{n_{i1}+\Delta,\mathbf{n}_{s+}}_{n_{i1},\mathbf{n}_{i+}})
    +\sum_{k>1}\sqrt{\frac{n_{sk}}{n_{i1}+1}}\mathcal{U}'_{sk,i1}(\prescript{\mathbf{m}_s}{\mathbf{m}_i}{}|\hat{U}'|^{n_{i1}+1+\Delta,\mathbf{n}_{s+}-\mathbf{e}_{sk}}_{n_{i1},\mathbf{n}_{i+}})\nonumber\\
    &\qquad\quad+\sum_k\sqrt{\frac{m_{ik}}{n_{i1}+1}}\mathcal{U}'_{ik,i1}(\prescript{\mathbf{m}_s}{\mathbf{m}_i-\mathbf{e}_{ik}}{}|\hat{U}'|^{n_{i1}+1+\Delta,\mathbf{n}_{s+}}_{n_{i1},\mathbf{n}_{i+}}),\nonumber\\
    =\,&C_UT\sum_k(\mathcal{U}'_{sk,i1}S_k+\mathcal{U}'_{ik,i1}I_k),
\end{align}
where we denote
\begin{align}
    S_1&=\sqrt{\frac{n_{i1}+1+\Delta}{n_{i1}+1}}\frac{(\prescript{\mathbf{m}_s}{\mathbf{m}_i}{}|\hat{U}'|^{n_{i1}+\Delta,\mathbf{n}_{s+}}_{n_{i1},\mathbf{n}_{i+}})}{C_UT}
    =\sqrt{\frac{n_{i1}+1+\Delta}{n_{i1}+1}}\sum_{n'=0}^{+\infty}\{\prescript{\mathbf{m}_s}{\mathbf{m}_i}{}|\hat{\mathcal{U}}|^{\Delta+n',\mathbf{n}_{s+}}_{n',\mathbf{n}_{i+}}\}\{\prescript{\Delta+n'}{n'}{}|\hat{\mathcal{B}}|^{n_{i1}+\Delta}_{n_{i1}}\},\nonumber\\
    S_k&=\sqrt{\frac{n_{sk}}{n_{i1}+1}}\frac{(\prescript{\mathbf{m}_s}{\mathbf{m}_i}{}|\hat{U}'|^{n_{i1}+1+\Delta,\mathbf{n}_{s+}-\mathbf{e}_{sk}}_{n_{i1},\mathbf{n}_{i+}})}{C_UT}
    =\sqrt{\frac{n_{sk}}{n_{i1}+1}}\sum_{n'=0}^{+\infty}\{\prescript{\mathbf{m}_s}{\mathbf{m}_i}{}|\hat{\mathcal{U}}|^{\Delta+n'+1,\mathbf{n}_{s+}-\mathbf{e}_{sk}}_{n',\mathbf{n}_{i+}}\}\{\prescript{\Delta+n'+1}{n'}{}|\hat{\mathcal{B}}|^{n_{i1}+1+\Delta}_{n_{i1}}\}\;\textrm{($k>1$)},\nonumber\\
    I_k&=\sqrt{\frac{m_{ik}}{n_{i1}+1}}\frac{(\prescript{\mathbf{m}_s}{\mathbf{m}_i-\mathbf{e}_{ik}}{}|\hat{U}'|^{n_{i1}+1+\Delta,\mathbf{n}_{s+}}_{n_{i1},\mathbf{n}_{i+}})}{C_UT}
    =\sqrt{\frac{m_{ik}}{n_{i1}+1}}\sum_{n'=0}^{+\infty}\{\prescript{\mathbf{m}_s}{\mathbf{m}_i-\mathbf{e}_{ik}}{}|\hat{\mathcal{U}}|^{\Delta+n'+1,\mathbf{n}_{s+}}_{n',\mathbf{n}_{i+}}\}\{\prescript{\Delta+n'+1}{n'}{}|\hat{\mathcal{B}}|^{n_{i1}+1+\Delta}_{n_{i1}}\}
\end{align}
for simplicity. We also have
\begin{align}\label{s23}
    \{\prescript{\Delta+n'}{n'}{}|\hat{\mathcal{B}}|^{n_{i1}+1+\Delta}_{n_{i1}+1}\}&=\frac{1}{\sqrt{n_{i1}+1}}\langle n_{i1}+1+\Delta;n'|\big(\hat{\mathcal{B}}\hat{a}^\dagger_{i1}\hat{\mathcal{B}}^\dagger\big)\hat{\mathcal{B}}|\Delta+n';n_{i1}\rangle\nonumber\\
    &=\sqrt{\frac{n_{i1}+1+\Delta}{n_{i1}+1}}R\{\prescript{\Delta+n'}{n'}{}|\hat{\mathcal{B}}|^{n_{i1}+\Delta}_{n_{i1}}\}+\sqrt{\frac{n'}{n_{i1}+1}}T\{\prescript{\Delta+n'}{n'-1}{}|\hat{\mathcal{B}}|^{n_{i1}+1+\Delta}_{n_{i1}}\},
\end{align}
\begin{align}\label{s24}
    \{\prescript{\Delta+n'+1}{n'}{}|\hat{\mathcal{B}}|^{n_{i1}+1+\Delta}_{n_{i1}}\}&=\frac{1}{\sqrt{\Delta+n'+1}}\langle n_{i1}+1+\Delta;n'|\big(\hat{\mathcal{B}}\hat{a}^\dagger_{s1}\hat{\mathcal{B}}^\dagger\big)\hat{\mathcal{B}}|\Delta+n';n_{i1}\rangle\nonumber\\
    &=\sqrt{\frac{n_{i1}+1+\Delta}{\Delta+n'+1}}T\{\prescript{\Delta+n'}{n'}{}|\hat{\mathcal{B}}|^{n_{i1}+\Delta}_{n_{i1}}\}-\sqrt{\frac{n'}{\Delta+n'+1}}R\{\prescript{\Delta+n'}{n'-1}{}|\hat{\mathcal{B}}|^{n_{i1}+1+\Delta}_{n_{i1}}\},
\end{align}
\begin{align}\label{s25}
    \{\prescript{\mathbf{m}_s}{\mathbf{m}_i}{}|\hat{\mathcal{U}}|^{\Delta+n',\mathbf{n}_{s+}}_{n',\mathbf{n}_{i+}}\}
    &=\frac{1}{\sqrt{n'}}\langle\Delta+n',\mathbf{n}_{s+};\mathbf{m}_i|\big(\hat{\mathcal{U}}\hat{a}^\dagger_{i1}\hat{\mathcal{U}}^\dagger\big)\hat{\mathcal{U}}|\mathbf{m}_s;n'-1,\mathbf{n}_{i+}\rangle\nonumber\\
    &=\sqrt{\frac{\Delta+n'}{n'}}\mathcal{U}_{s1,i1}\{\prescript{\mathbf{m}_s}{\mathbf{m}_i}{}|\hat{\mathcal{U}}|^{\Delta+n'-1,\mathbf{n}_{s+}}_{n'-1,\mathbf{n}_{i+}}\}
    +\sum_{k>1}\sqrt{\frac{n_{sk}}{n'}}\mathcal{U}_{sk,i1}\{\prescript{\mathbf{m}_s}{\mathbf{m}_i}{}|\hat{\mathcal{U}}|^{\Delta+n',\mathbf{n}_{s+}-\mathbf{e}_{sk}}_{n'-1,\mathbf{n}_{i+}}\}\nonumber\\
    &\qquad\qquad+\sum_{k}\sqrt{\frac{m_{ik}}{n'}}\mathcal{U}_{ik,i1}\{\prescript{\mathbf{m}_s}{\mathbf{m}_i-\mathbf{e}_{ik}}{}|\hat{\mathcal{U}}|^{\Delta+n',\mathbf{n}_{s+}}_{n'-1,\mathbf{n}_{i+}}\}\;\textrm{($n'\geq1$)},
\end{align}
[the second terms in Eqs.\ \eqref{s23} and \eqref{s24} vanish if $n'=0$] and
\begin{align}\label{s26}
    &\sum_{n'=0}^{+\infty}\{\prescript{\mathbf{m}_s}{\mathbf{m}_i}{}|\hat{\mathcal{U}}|^{\Delta+n',\mathbf{n}_{s+}}_{n',\mathbf{n}_{i+}}\}\overbrace{\{\prescript{\Delta+n'}{n'}{}|\hat{\mathcal{B}}|^{n_{i1}+1+\Delta}_{n_{i1}+1}\}}^\textrm{Apply Eq.\ \eqref{s23}}\nonumber\\
    =\,&\sum_{n'=0}^{+\infty}\{\prescript{\mathbf{m}_s}{\mathbf{m}_i}{}|\hat{\mathcal{U}}|^{\Delta+n',\mathbf{n}_{s+}}_{n',\mathbf{n}_{i+}}\}\left(\sqrt{\frac{n_{i1}+1+\Delta}{n_{i1}+1}}R\{\prescript{\Delta+n'}{n'}{}|\hat{\mathcal{B}}|^{n_{i1}+\Delta}_{n_{i1}}\}+\sqrt{\frac{n'}{n_{i1}+1}}T\{\prescript{\Delta+n'}{n'-1}{}|\hat{\mathcal{B}}|^{n_{i1}+1+\Delta}_{n_{i1}}\}\right)\nonumber\\
    =\,&RS_1+T\overbrace{\sum_{n'=1}^{+\infty}\sqrt{\frac{n'}{n_{i1}+1}}\{\prescript{\mathbf{m}_s}{\mathbf{m}_i}{}|\hat{\mathcal{U}}|^{\Delta+n',\mathbf{n}_{s+}}_{n',\mathbf{n}_{i+}}\}\{\prescript{\Delta+n'}{n'-1}{}|\hat{\mathcal{B}}|^{n_{i1}+1+\Delta}_{n_{i1}}\}}^\textrm{Let it be $O$}.
\end{align}
From Eqs.\ \eqref{s24} and \eqref{s25},
\begin{align}
    O&=\mathcal{U}_{s1,i1}\sum_{n'=0}^{+\infty}\sqrt{\frac{\Delta+n'+1}{n_{i1}+1}}\{\prescript{\mathbf{m}_s}{\mathbf{m}_i}{}|\hat{\mathcal{U}}|^{\Delta+n',\mathbf{n}_{s+}}_{n',\mathbf{n}_{i+}}\}\{\prescript{\Delta+n'+1}{n'}{}|\hat{\mathcal{B}}|^{n_{i1}+1+\Delta}_{n_{i1}}\}+\sum_{k>1}\mathcal{U}_{sk,i1}S_k+\sum_k\mathcal{U}_{ik,i1}I_k\nonumber\\
    &=T\mathcal{U}_{s1,i1}S_1+\sum_{k>1}\mathcal{U}_{sk,i1}S_k+\sum_k\mathcal{U}_{ik,i1}I_k-R\mathcal{U}_{s1,i1}O,
\end{align}
where the substitution $n'\to n'+1$ is made to obtain the first line, so
\begin{equation}
    O=\frac{T\mathcal{U}_{s1,i1}}{1+R\mathcal{U}_{s1,i1}}S_1+\sum_{k>1}\frac{\mathcal{U}_{sk,i1}}{1+R\mathcal{U}_{s1,i1}}S_k+\sum_k\frac{\mathcal{U}_{ik,i1}}{1+R\mathcal{U}_{s1,i1}}I_k
\end{equation}
and Eq.\ \eqref{s26} becomes
\begin{equation}
    \left(R+\frac{T^2\mathcal{U}_{s1,i1}}{1+R\mathcal{U}_{s1,i1}}\right)S_1+\sum_{k>1}\frac{T\mathcal{U}_{sk,i1}}{1+R\mathcal{U}_{s1,i1}}S_k+\sum_k\frac{T\mathcal{U}_{ik,i1}}{1+R\mathcal{U}_{s1,i1}}I_k=\sum_k(\mathcal{U}'_{sk,i1}S_k+\mathcal{U}'_{ik,i1}I_k),
\end{equation}
which means Eq.\ \eqref{s21} equals
\begin{equation}
    C_UT\sum_{n'=0}^{+\infty}\{\prescript{\mathbf{m}_s}{\mathbf{m}_i}{}|\hat{\mathcal{U}}|^{\Delta+n',\mathbf{n}_{s+}}_{n',\mathbf{n}_{i+}}\}\{\prescript{\Delta+n'}{n'}{}|\hat{\mathcal{B}}|^{n_{i1}+1+\Delta}_{n_{i1}+1}\}=(\prescript{\mathbf{m}_s}{\mathbf{m}_i}{}|\hat{U}'|^{n_{i1}+1+\Delta,\mathbf{n}_{s+}}_{n_{i1}+1,\mathbf{n}_{i+}}),
\end{equation}
proving this lemma.

\end{widetext}

\section{Duality with general state detection}\label{geneproj}

From Eq.\ \eqref{purestateposts}, by inserting a complete \emph{s} basis, we have the duality theorem for the \emph{s} state after \emph{i} postselection
\begin{align}
    \langle\psi_i|\hat{U}|\psi_{s0};\psi_{i0}\rangle&=\sum_{\mathbf{n}_s}|\mathbf{n}_s\rangle\langle\mathbf{n}_s;\psi_i|\hat{U}|\psi_{s0};\psi_{i0}\rangle\nonumber\\
    &=C_U\sum_{\mathbf{n}_s}|\mathbf{n}_s\rangle\langle\mathbf{n}_s;\psi_{i0}^\ast|\hat{\mathcal{U}}|\psi_{s0};\psi_i^\ast\rangle\nonumber\\
    &=C_U\langle\psi_{i0}^\ast|\hat{\mathcal{U}}|\psi_{s0};\psi_i^\ast\rangle.
\end{align}

The expectation value of a separable operator $\hat{O}_s\otimes\hat{O}_i$ after the nonlinear system
\begin{align}
    &\langle\psi_{s0};\psi_{i0}|\hat{U}^\dagger\big(\hat{O}_s\otimes\hat{O}_i\big)\hat{U}|\psi_{s0};\psi_{i0}\rangle\nonumber\\
    =\,&\sum_{\mathbf{m}_s,\mathbf{m}_i,\mathbf{n}_s,\mathbf{n}_i}
    \langle\psi_{s0};\psi_{i0}|\hat{U}^\dagger|\mathbf{m}_s;\mathbf{m}_i\rangle\langle\mathbf{m}_s|\hat{O}_s|\mathbf{n}_s\rangle\langle\mathbf{m}_i|\hat{O}_i|\mathbf{n}_i\rangle\nonumber\\
    &\qquad\qquad\quad\times\langle\mathbf{n}_s;\mathbf{n}_i|\hat{U}|\psi_{s0};\psi_{i0}\rangle\nonumber\\
    =\,&|C_U|^2\sum_{\mathbf{m}_s,\mathbf{m}_i,\mathbf{n}_s,\mathbf{n}_i}\langle\mathbf{n}_s;\psi_{i0}^\ast|\hat{\mathcal{U}}|\psi_{s0};\mathbf{n}_i\rangle\langle\mathbf{n}_i|\hat{O}_i^\mathrm{T}|\mathbf{m}_i\rangle
    \nonumber\\
    &\qquad\qquad\qquad\quad\times\langle\psi_{s0};\mathbf{m}_i|\hat{\mathcal{U}}^\dagger|\mathbf{m}_s;\psi_{i0}^\ast\rangle\langle\mathbf{m}_s|\hat{O}_s|\mathbf{n}_s\rangle\nonumber\\
    =\,&|C_U|^2\sum_{\mathbf{n}_s}\langle\mathbf{n}_s;\psi_{i0}^\ast|\hat{\mathcal{U}}\big(|\psi_{s0}\rangle\langle\psi_{s0}|\otimes\hat{O}_i^\mathrm{T}\big)\hat{\mathcal{U}}^\dagger|\psi_{i0}^\ast\rangle\hat{O}_s|\mathbf{n}_s\rangle\nonumber\\
    =\,&|C_U|^2\operatorname{Tr}\big[\langle\psi_{i0}^\ast|\hat{\mathcal{U}}\big(|\psi_{s0}\rangle\langle\psi_{s0}|\otimes\hat{O}_i^\mathrm{T}\big)\hat{\mathcal{U}}^\dagger|\psi_{i0}^\ast\rangle\hat{O}_s\big].
\end{align}
If $\hat{O}_s$ and $\hat{O}_i$ are Hermitian and positive-definite, apart from $|C_U|^2$, this is the expectation value of $\hat{O}_s$ with respect to the \emph{s} state from the unnormalized initial state $|\psi_{s0}\rangle\langle\psi_{s0}|\otimes\hat{O}_i^\mathrm{T}$ after the PTR setup $\hat{\mathcal{U}}$ and \emph{i} postselection to $|\psi_{i0}^\ast\rangle$. If $\hat{O}_s=|\psi_s\rangle\langle\psi_s|$ and $\hat{O}_i=|\psi_i\rangle\langle\psi_i|$, the expectation value is the postselection probability. If some \emph{s} or \emph{i} paths are undetected, identity operators are used for direct product to obtain $\hat{O}_s$ or $\hat{O}_i$. Furthermore, $|\psi_{s0}\rangle$ and $|\psi_{i0}\rangle$ can be generalized to operators \cite{pnas.2010827117}.

If linear losses are present in the nonlinear system, we can analyze an lossless one equivalently by replacing each absorptive media with a BS coupled to an undetected auxiliary path with the vacuum input. However, the calculation involves summing the probabilities of postselecting the undetected paths to all possible photon numbers, which is generally not simple.

\section{Case of polychromatic down-converted lights}\label{polychrom}

For simplicity, the pump beam is monochromatic at $\omega_p$. Different light frequencies can be treated as different groups of spatial paths. We use $\hat{a}^\dagger_{sj,\omega}$ to denote the creation operator on path $sj$ with the frequency $\omega$. Linear system operators only act in a single frequency group, denoted as $\hat{L}_{s,\omega_s}$ and $\hat{L}_{i,\omega_i}$. If there is a PDC device, PDC operations act on an \emph{s} and an \emph{i} path of all pairs of groups with $\omega_s$ and $\omega_i=\omega_p-\omega_s$, which means different pairs are isolated. We let the \emph{s} and \emph{i} frequency ranges (with no overlap; their widths are the same) be $D_s$ and $D_i$, so, if $\omega_s\notin D_s$, there is no PDC \cite{polyfilter}. The scattering matrix of the PTR setup depends on $\omega_s$ or $\omega_i$ and can be written as $\boldsymbol{\mathcal{U}}_{\omega_s}$.

When a detector at path $ij$ records one photon at the time $t_0$, the postselected state of path $ij$ (in all frequency groups) is $\int_0^{+\infty}d\omega e^{i\omega t_0}\hat{a}^\dagger_{ij,\omega}|\mathrm{vac}\rangle$. From the duality, the initial \emph{i} state of the PTR setup is its conjugate $\int_0^{+\infty}d\omega e^{-i\omega t_0}\hat{a}^\dagger_{ij,\omega}|\mathrm{vac}\rangle$. Ignoring groups with $\omega_i\notin D_i$ , the effective initial state is $\int_{D_i}d\omega e^{-i\omega t_0}\hat{a}^\dagger_{ij,\omega}|\mathrm{vac}\rangle$, which is, however, pulsed at $-t_0$. If the initial states of the nonlinear system are vacuum, after the PTR setup and vacuum \emph{i} postseletion, the \emph{s} state is
\begin{equation}
    \int_{D_s}d\omega e^{-i(\omega_p-\omega) t_0}\sum_k(\mathcal{U}_{\omega})_{sk,ij}\hat{a}^\dagger_{sk,\omega}|\mathrm{vac}\rangle.
\end{equation}

Then, we consider one PDC device with a constant gain, followed by propagating the distances $d+\Delta d$ and $d$ in the vacuum for the \emph{s} and \emph{i} lights respectively. A step-by-step analysis of the PTR setup is that: after the first propagation, the \emph{i} state $\int_{D_i}d\omega e^{i\omega(d/c-t_0)}\hat{a}^\dagger_{ij,\omega}|\mathrm{vac}\rangle$ is pulsed at $d/c-t_0$ ($c$ is the speed of light in vacuum); after the reflection (with frequency inversion), the \emph{s} state $R\int_{D_s}d\omega e^{i(\omega_p-\omega)(d/c-t_0)}\hat{a}^\dagger_{sj,\omega}|\mathrm{vac}\rangle$ is at $t_0-d/c$; after the second propagation, it is at $t_0+\Delta d/c$, when the \emph{s} light should be detected in the nonlinear system. If minus signs are taken for the \emph{i} pulse times in the PTR setup, they are $t_0$ at the \emph{i} detector and $t_0-d/c$ at the hypothetical BS, which means the \emph{i} light travels backward in time, and the arrow of time switches when being reflected by hypothetical BSs.

If the same PDC device is added after the propagations to form an SU(1,1) interferometer, from Eq.\ \eqref{su11ptr}, $(\mathcal{U}_{\omega})_{si}=2R\cos(\Phi/2)e^{i\Phi/2}/(1+R^2e^{i\Phi})$, where $\Phi=(\omega_p d+\omega\Delta d)/c$, so the \emph{s} state is
\begin{equation}
    e^{-i\omega_p t_0}\int_{D_s}d\omega e^{i\omega t_0}\frac{R(1+e^{i\Phi})}{1+R^2e^{i\Phi}}\hat{a}^\dagger_{s,\omega}|\mathrm{vac}\rangle.
\end{equation}
Ignoring the global phase, compared to a single pulse (with a finite width) at $t_0$ whose state is $\int_{D_s}d\omega e^{i\omega t_0}\hat{a}^\dagger_{s,\omega}|\mathrm{vac}\rangle$, the extra term
\begin{equation}
    \frac{R(1+e^{i\Phi})}{1+R^2e^{i\Phi}}=R\left[1+T^2\sum_{n=1}^{+\infty}(-R^2)^{n-1}e^{in\Phi}\right],
\end{equation}
whose Fourier transform is
\begin{equation}
    R\delta(t)+RT^2\sum_{n=1}^{+\infty}(-R^2)^{n-1}e^{in\omega_p d_i/c}\delta\left(t-\frac{n\Delta d}{c}\right).
\end{equation}
From the convolution theorem, the temporal state of the \emph{s} photon is a decaying train of pulses with the interval $\Delta d/c$ (if $\Delta d=0$, it is still a single pulse), while a realistic FP cavity has an interval $(2d+\Delta d)/c$.

\section{Gaussian operation approach to the duality}\label{gaussduality}

\subsection{Multipath transfer matrix}

In classical optics, when analyzing multilayer thin-film structures, the transfer matrix is used to connect the forward- and backward-propagating light amplitudes at one side of the interface to the other, while the scattering matrix connecting the output amplitudes to the input ones follows the causality. The transfer matrix of cascaded BSs forming a cavity is simply the product of individual matrices \cite{quanphoton}, unlike the Redheffer star product for scattering matrices.

In the multipath case, considering the PTR setup, the transfer matrix is written as
\begin{equation}
    \boldsymbol{T}=\begin{pmatrix}\boldsymbol{T}_{ss}&\boldsymbol{T}_{si}\\\boldsymbol{T}_{is}&\boldsymbol{T}_{ii}\end{pmatrix}.
\end{equation}
If the \emph{s} (forward) and \emph{i} (backward) classical amplitudes on the left side of this setup are $\boldsymbol{\alpha}_s$ and $\boldsymbol{\alpha}'_i$, and those on the right side are $\boldsymbol{\alpha}'_s$ and $\boldsymbol{\alpha}_i$, their relations are
\begin{equation}
    \begin{pmatrix}\boldsymbol{\alpha}'_s\\\boldsymbol{\alpha}_i\end{pmatrix}=\boldsymbol{T}\begin{pmatrix}\boldsymbol{\alpha}_s\\\boldsymbol{\alpha}'_i\end{pmatrix},\;
    \begin{pmatrix}\boldsymbol{\alpha}'_s\\\boldsymbol{\alpha}'_i\end{pmatrix}=\boldsymbol{\mathcal{U}}\begin{pmatrix}\boldsymbol{\alpha}_s\\\boldsymbol{\alpha}_i\end{pmatrix}.
\end{equation}
By solving $(\boldsymbol{\alpha}'_s,\boldsymbol{\alpha}'_i)$ and $(\boldsymbol{\alpha}'_s,\boldsymbol{\alpha}_i)$ respectively, we have
\begin{gather}
    \boldsymbol{\mathcal{U}}=\begin{pmatrix}
    \boldsymbol{T}_{ss}-\boldsymbol{T}_{si}\boldsymbol{T}_{ii}^{-1}\boldsymbol{T}_{is}&\boldsymbol{T}_{si}\boldsymbol{T}_{ii}^{-1}\\
    -\boldsymbol{T}_{ii}^{-1}\boldsymbol{T}_{is}&\boldsymbol{T}_{ii}^{-1}\end{pmatrix},\nonumber\\
    \boldsymbol{T}=\begin{pmatrix}
    \boldsymbol{\mathcal{U}}_{ss}-\boldsymbol{\mathcal{U}}_{si}\boldsymbol{\mathcal{U}}_{ii}^{-1}\boldsymbol{\mathcal{U}}_{is}&\boldsymbol{\mathcal{U}}_{si}\boldsymbol{\mathcal{U}}_{ii}^{-1}\\
    -\boldsymbol{\mathcal{U}}_{ii}^{-1}\boldsymbol{\mathcal{U}}_{is}&\boldsymbol{\mathcal{U}}_{ii}^{-1}\end{pmatrix}.
\end{gather}
One can verify the scattering matrix of a cascaded system $\boldsymbol{T}_2\boldsymbol{T}_1$ is $\boldsymbol{\mathcal{U}}_1\star\boldsymbol{\mathcal{U}}_2$.

For a BS coupling paths $s1$ and $i1$,
\begin{gather}
    \boldsymbol{T}_{ss}=\boldsymbol{T}_{ii}=\operatorname{diag}\{1/T,1,1,\ldots\}=\operatorname{diag}\{\cosh r,1,1,\ldots\},\nonumber\\
    \boldsymbol{T}_{si}=\boldsymbol{T}_{is}=\operatorname{diag}\{R/T,0,0,\ldots\}=\operatorname{diag}\{\sinh r,0,0,\ldots\}.
\end{gather}
For two linear lossless systems $\boldsymbol{L}_s$ and $\boldsymbol{L}_i$, $\boldsymbol{T}_{si}$ and $\boldsymbol{T}_{is}$ are zero matrices, $\boldsymbol{T}_{ss}=\boldsymbol{L}_s$, and $\boldsymbol{T}_{ii}=\boldsymbol{L}_i^\ast$ [because the \emph{i} light goes through the system $\boldsymbol{L}_i^\mathrm{T}$ and becomes the amplitudes on the left side, so the field on the right side is obtained by multiplying its inverse $(\boldsymbol{L}_i^\mathrm{T})^{-1}=(\boldsymbol{L}_i^\mathrm{T})^\dagger=\boldsymbol{L}_i^\ast$]. For a linear setup $\boldsymbol{T}$ which is a cascade of hypothetical BSs and linear system pairs, letting $\boldsymbol{Z}=\begin{psmallmatrix}\boldsymbol{I}&\\&-\boldsymbol{I}\end{psmallmatrix}$, a complex form of the symplectic condition \cite{matrix} $\boldsymbol{T}\boldsymbol{Z}\boldsymbol{T}^\dagger=\boldsymbol{Z}$ holds for both $\boldsymbol{T}$ and $\boldsymbol{T}^\dagger$ (i.e., $\boldsymbol{T}^\dagger\boldsymbol{Z}\boldsymbol{T}=\boldsymbol{Z}$) \cite{verifyt}. We have $|\det\boldsymbol{T}|=1$ and the following.

\textit{Lemma 3}. For a transfer matrix $\boldsymbol{T}$ of a PTR setup,
\begin{equation}
    \left(\frac{\boldsymbol{T}\boldsymbol{T}^\dagger+\boldsymbol{I}_2}{2}\right)^{-1}=\boldsymbol{V}^\dagger\boldsymbol{V},
\end{equation}
where $\boldsymbol{I}_2=\begin{psmallmatrix}\boldsymbol{I}&\\&\boldsymbol{I}\end{psmallmatrix}$ and
\begin{equation}
    \boldsymbol{V}=\begin{pmatrix}\boldsymbol{I}&-\boldsymbol{T}_{si}\boldsymbol{T}_{ii}^{-1}\\
    &-\boldsymbol{T}_{ii}^{-1}\end{pmatrix}.
\end{equation}

\textit{Proof}. From the conditions $\boldsymbol{T}\boldsymbol{Z}\boldsymbol{T}^\dagger=\boldsymbol{Z}$ and $\boldsymbol{T}^\dagger\boldsymbol{Z}\boldsymbol{T}=\boldsymbol{Z}$, we have
\begin{align}\label{symproperties}
    \boldsymbol{T}_{ss}\boldsymbol{T}_{ss}^\dagger-\boldsymbol{T}_{si}\boldsymbol{T}_{si}^\dagger=\boldsymbol{I},&\;
    \boldsymbol{T}_{ss}^\dagger\boldsymbol{T}_{ss}-\boldsymbol{T}_{is}^\dagger\boldsymbol{T}_{is}=\boldsymbol{I},\nonumber\\
    \boldsymbol{T}_{ii}\boldsymbol{T}_{ii}^\dagger-\boldsymbol{T}_{is}\boldsymbol{T}_{is}^\dagger=\boldsymbol{I},&\;
    \boldsymbol{T}_{ii}^\dagger\boldsymbol{T}_{ii}-\boldsymbol{T}_{si}^\dagger\boldsymbol{T}_{si}=\boldsymbol{I},\nonumber\\
    \boldsymbol{T}_{ss}\boldsymbol{T}_{is}^\dagger=\boldsymbol{T}_{si}\boldsymbol{T}_{ii}^\dagger,&\;
    \boldsymbol{T}_{ss}^\dagger\boldsymbol{T}_{si}=\boldsymbol{T}_{is}^\dagger\boldsymbol{T}_{ii},\nonumber\\
    \boldsymbol{T}_{is}\boldsymbol{T}_{ss}^\dagger=\boldsymbol{T}_{ii}\boldsymbol{T}_{si}^\dagger,&\;
    \boldsymbol{T}_{si}^\dagger\boldsymbol{T}_{ss}=\boldsymbol{T}_{ii}^\dagger\boldsymbol{T}_{is},
\end{align}
so $(\boldsymbol{T}_{ii}^\dagger)^{-1}\boldsymbol{T}_{si}^\dagger=\boldsymbol{T}_{is}\boldsymbol{T}_{ss}^{-1}$. Because $(\boldsymbol{A}^{-1})^\dagger=(\boldsymbol{A}^\dagger)^{-1}$,
\begin{align}
    \boldsymbol{V}^\dagger\boldsymbol{V}&=\begin{pmatrix}\boldsymbol{I}&\\
    -(\boldsymbol{T}_{ii}^\dagger)^{-1}\boldsymbol{T}_{si}^\dagger&-(\boldsymbol{T}_{ii}^\dagger)^{-1}\end{pmatrix}
    \begin{pmatrix}\boldsymbol{I}&-\boldsymbol{T}_{si}\boldsymbol{T}_{ii}^{-1}\\
    &-\boldsymbol{T}_{ii}^{-1}\end{pmatrix}\nonumber\\
    &=\begin{pmatrix}\boldsymbol{I}&-\boldsymbol{T}_{si}\boldsymbol{T}_{ii}^{-1}\\
    -(\boldsymbol{T}_{ii}^\dagger)^{-1}\boldsymbol{T}_{si}^\dagger&(\boldsymbol{T}_{ii}^\dagger)^{-1}(\boldsymbol{I}+\boldsymbol{T}_{si}^\dagger\boldsymbol{T}_{si})\boldsymbol{T}_{ii}^{-1}\end{pmatrix}\nonumber\\
    &=\begin{pmatrix}\boldsymbol{I}&-\boldsymbol{T}_{si}\boldsymbol{T}_{ii}^{-1}\\
    -\boldsymbol{T}_{is}\boldsymbol{T}_{ss}^{-1}&\boldsymbol{I}\end{pmatrix}.
\end{align}
Also,
\begin{equation}
    \frac{\boldsymbol{T}\boldsymbol{T}^\dagger+\boldsymbol{I}_2}{2}=
    \begin{pmatrix}\boldsymbol{T}_{ss}\boldsymbol{T}_{ss}^\dagger&\boldsymbol{T}_{ss}\boldsymbol{T}_{is}^\dagger\\\boldsymbol{T}_{ii}\boldsymbol{T}_{si}^\dagger&\boldsymbol{T}_{ii}\boldsymbol{T}_{ii}^\dagger\end{pmatrix}
    =\begin{pmatrix}\boldsymbol{T}_{ss}&\\&\boldsymbol{T}_{ii}\end{pmatrix}\boldsymbol{T}^\dagger
\end{equation}
and
\begin{equation}
    (\boldsymbol{T}^\dagger)^{-1}=\begin{pmatrix}\boldsymbol{T}_{ss}&-\boldsymbol{T}_{si}\\-\boldsymbol{T}_{is}&\boldsymbol{T}_{ii}\end{pmatrix}
\end{equation}
(which can be verified by multiplying $\boldsymbol{T}^\dagger$), so
\begin{equation}
    \left(\frac{\boldsymbol{T}\boldsymbol{T}^\dagger+\boldsymbol{I}_2}{2}\right)^{-1}
    =(\boldsymbol{T}^\dagger)^{-1}\begin{pmatrix}\boldsymbol{T}_{ss}^{-1}&\\&\boldsymbol{T}_{ii}^{-1}\end{pmatrix}=\boldsymbol{V}^\dagger\boldsymbol{V}.
\end{equation}

\subsection{Gaussian operation and \emph{Q}-function relation}

A nonlinear system with nondegenerate PDCs and linear lossless systems performs a Gaussian operation to the quantum optical field. Such an operation can be described by a symplectic matrix $\boldsymbol{F}$ which determines the Bogoliubov transformation to the creation and annihilation operators. Using the $\boldsymbol{\alpha},\boldsymbol{\alpha}^\ast$ basis (there are other notations in the literature) which becomes $\boldsymbol{\alpha}_s,\boldsymbol{\alpha}_i,\boldsymbol{\alpha}_s^\ast,\boldsymbol{\alpha}_i^\ast$ (they are column vectors) in our scenario, for a PDC or a linear system pair, the symplectic matrix is related to its transfer matrix
\begin{equation}\label{Fpdc}
    \boldsymbol{F}=\begin{pmatrix}
        \boldsymbol{T}_{ss}&&&\boldsymbol{T}_{si}\\
        &\boldsymbol{T}_{ii}^\ast&\boldsymbol{T}_{is}^\ast&\\
        &\boldsymbol{T}_{si}^\ast&\boldsymbol{T}_{ss}^\ast&\\
        \boldsymbol{T}_{is}&&&\boldsymbol{T}_{ii}
    \end{pmatrix}.
\end{equation}
If two systems satisfy this relation, by calculating $\boldsymbol{F}_2\boldsymbol{F}_1$ and $\boldsymbol{T}_2\boldsymbol{T}_1$, the whole system also does. The empty blocks in $\boldsymbol{F}$ are due to the absence of single-mode squeezing.

A multimode Gaussian state is described by the covariance matrix $\boldsymbol{\sigma}$ and the displacement vector
\begin{equation}
    \bar{\mathbf{a}}=\begin{pmatrix}
        \boldsymbol{\alpha}_s\\
        \boldsymbol{\alpha}_i\\
        \boldsymbol{\alpha}_s^\ast\\
        \boldsymbol{\alpha}_i^\ast
    \end{pmatrix}=(\boldsymbol{\alpha}_s^\mathrm{T},\boldsymbol{\alpha}_i^\mathrm{T},\boldsymbol{\alpha}_s^\dagger,\boldsymbol{\alpha}_i^\dagger)^\mathrm{T}.
\end{equation}
For a multimode coherent (including vacuum) state, $\boldsymbol{\sigma}=\boldsymbol{I}_4/2$, where $\boldsymbol{I}_4=\begin{psmallmatrix}\boldsymbol{I}_2&\\&\boldsymbol{I}_2\end{psmallmatrix}$, and $\boldsymbol{\alpha}_s,\boldsymbol{\alpha}_i$ are indeed the displacement parameter. After the nonlinear system $\boldsymbol{F}$, they become $\boldsymbol{\sigma}'=\boldsymbol{F}\boldsymbol{\sigma}\boldsymbol{F}^\dagger$ and $\bar{\mathbf{a}}'=\boldsymbol{F}\bar{\mathbf{a}}$, respectively. Letting $\mathbf{b}=(\boldsymbol{\beta}_s^\mathrm{T},\boldsymbol{\beta}_i^\mathrm{T},\boldsymbol{\beta}_s^\dagger,\boldsymbol{\beta}_i^\dagger)^\mathrm{T}$ and $\boldsymbol{\sigma}'_Q=\boldsymbol{\sigma}'+\boldsymbol{I}_4/2$, ignoring the term of a power of $\pi$, the \emph{Q}-function of the final state
\begin{equation}\label{gaussq}
    Q(\boldsymbol{\beta}_s,\boldsymbol{\beta}_i)\propto\sqrt{\det(\boldsymbol{\sigma}'_Q)^{-1}}e^{-\frac{1}{2}(\mathbf{b}-\bar{\mathbf{a}}')^\dagger(\boldsymbol{\sigma}'_Q)^{-1}(\mathbf{b}-\bar{\mathbf{a}}')}.
\end{equation}

For simplicity, we permute the second and fourth blocks of the vectors so that $\bar{\mathbf{a}}=(\boldsymbol{\alpha}_s^\mathrm{T},\boldsymbol{\alpha}_i^\dagger,\boldsymbol{\alpha}_s^\dagger,\boldsymbol{\alpha}_i^\mathrm{T})^\mathrm{T}$ and $\mathbf{b}=(\boldsymbol{\beta}_s^\mathrm{T},\boldsymbol{\beta}_i^\dagger,\boldsymbol{\beta}_s^\dagger,\boldsymbol{\beta}_i^\mathrm{T})^\mathrm{T}$. Then, the new form of $\boldsymbol{F}$ is $\begin{psmallmatrix}\boldsymbol{T}&\\&\boldsymbol{T}^\ast\end{psmallmatrix}$. The permutation time of square matrices must be even, so $\det\boldsymbol{\sigma}_Q$ is unchanged. If the initial $\boldsymbol{\sigma}=\boldsymbol{I}_4/2$,
\begin{align}
    (\boldsymbol{\sigma}'_Q)^{-1}&=\left(\frac{\boldsymbol{F}\boldsymbol{F}^\dagger+\boldsymbol{I}_4}{2}\right)^{-1}=\begin{pmatrix}
        \boldsymbol{V}^\dagger\boldsymbol{V}&\\&\boldsymbol{V}^\mathrm{T}\boldsymbol{V}^\ast
    \end{pmatrix}\nonumber\\
    &=\begin{pmatrix}
        \boldsymbol{V}&\\&\boldsymbol{V}^\ast
    \end{pmatrix}^\dagger\begin{pmatrix}
        \boldsymbol{V}&\\&\boldsymbol{V}^\ast
    \end{pmatrix},
\end{align}
\begin{equation}
    \sqrt{\det(\boldsymbol{\sigma}'_Q)^{-1}}=|\det\boldsymbol{V}|^2=|\det\boldsymbol{T}_{ii}^{-1}|^2=|\det\boldsymbol{\mathcal{U}}_{ii}|^2,
\end{equation}
and the exponent in Eq.\ \eqref{gaussq} equals
\begin{align}
    -\frac{1}{2}\left|\begin{pmatrix}
        \boldsymbol{V}&\\&\boldsymbol{V}^\ast
    \end{pmatrix}(\mathbf{b}-\boldsymbol{F}\bar{\mathbf{a}})\right|^2
    &=-\left|\boldsymbol{V}\left[\begin{pmatrix}\boldsymbol{\beta}_s\\\boldsymbol{\beta}_i^\ast\end{pmatrix}-\boldsymbol{T}\begin{pmatrix}\boldsymbol{\alpha}_s\\\boldsymbol{\alpha}_i^\ast\end{pmatrix}\right]\right|^2\nonumber\\
    &=-\left|\begin{pmatrix}\boldsymbol{\beta}_s\\\boldsymbol{\alpha}_i^\ast\end{pmatrix}-\boldsymbol{\mathcal{U}}\begin{pmatrix}\boldsymbol{\alpha}_s\\\boldsymbol{\beta}_i^\ast\end{pmatrix}\right|^2,
\end{align}
the same as the exponent in the \emph{Q}-function $Q_\mathrm{PTR}(\boldsymbol{\beta}_s,\boldsymbol{\alpha}_i^\ast)$ from the PTR setup with the initial multimode coherent states $\boldsymbol{\alpha}_s$ and $\boldsymbol{\beta}_i^\ast$ [the final covariance matrix is still $\boldsymbol{I}_4/2$, so $\det(\boldsymbol{\sigma}'_Q)^{-1}=1$ in this case]. So, the \emph{Q}-function relation is $Q(\boldsymbol{\beta}_s,\boldsymbol{\beta}_i)=|C_U|^2Q_\mathrm{PTR}(\boldsymbol{\beta}_s,\boldsymbol{\alpha}_i^\ast)$ or, explicitly,
\begin{equation}
    \big|\langle\boldsymbol{\beta}_s;\boldsymbol{\beta}_i|\hat{U}|\boldsymbol{\alpha}_s;\boldsymbol{\alpha}_i\rangle\big|^2=\big|C_U\langle\boldsymbol{\beta}_s;\boldsymbol{\alpha}_i^\ast|\hat{\mathcal{U}}|\boldsymbol{\alpha}_s;\boldsymbol{\beta}_i^\ast\rangle\big|^2,
\end{equation}
where $|C_U|=|\det\boldsymbol{\mathcal{U}}_{ii}|$.

\subsection{Nonlinear-linear duality}

Letting the \emph{P}-functions of four multipath pure states be $P_{s0}(\boldsymbol{\alpha})$, $P_s(\boldsymbol{\alpha})$, $P_{i0}(\boldsymbol{\alpha})$, and $P_i(\boldsymbol{\alpha})$ (for example, $|\psi_{s0}\rangle\langle\psi_{s0}|=\int d\boldsymbol{\alpha} P_{s0}(\boldsymbol{\alpha})|\boldsymbol{\alpha}\rangle\langle\boldsymbol{\alpha}|$), because $|\psi_i^\ast\rangle\langle\psi_i^\ast|=\int d\boldsymbol{\alpha}P_i(\boldsymbol{\alpha})|\boldsymbol{\alpha}^\ast\rangle\langle\boldsymbol{\alpha}^\ast|$, we have
\begin{align}
    &\big|\langle\psi_s;\psi_i|\hat{U}|\psi_{s0};\psi_{i0}\rangle\big|^2\nonumber\\
    =\,&\int d\boldsymbol{\alpha}_s d\boldsymbol{\alpha}_i d\boldsymbol{\beta}_s d\boldsymbol{\beta}_i
    P_{s0}(\boldsymbol{\alpha}_s)P_{i0}(\boldsymbol{\alpha}_i)P_s(\boldsymbol{\beta}_s)P_i(\boldsymbol{\beta}_i)\nonumber\\
    &\quad\times\langle\boldsymbol{\beta}_s;\boldsymbol{\beta}_i|\hat{U}|\boldsymbol{\alpha}_s;\boldsymbol{\alpha}_i\rangle\langle\boldsymbol{\alpha}_s;\boldsymbol{\alpha}_i|\hat{U}^\dagger|\boldsymbol{\beta}_s;\boldsymbol{\beta}_i\rangle\nonumber\\
    =\,&|C_U|^2\int d\boldsymbol{\alpha}_s d\boldsymbol{\alpha}_i d\boldsymbol{\beta}_s d\boldsymbol{\beta}_i
    P_{s0}(\boldsymbol{\alpha}_s)P_{i0}(\boldsymbol{\alpha}_i)P_s(\boldsymbol{\beta}_s)P_i(\boldsymbol{\beta}_i)\nonumber\\
    &\qquad\quad\;\times\langle\boldsymbol{\beta}_s;\boldsymbol{\alpha}_i^\ast|\hat{\mathcal{U}}|\boldsymbol{\alpha}_s;\boldsymbol{\beta}_i^\ast\rangle
    \langle\boldsymbol{\alpha}_s;\boldsymbol{\beta}_i^\ast|\hat{\mathcal{U}}^\dagger|\boldsymbol{\beta}_s;\boldsymbol{\alpha}_i^\ast\rangle\nonumber\\
    =\,&\big|C_U\langle\psi_s;\psi_{i0}^\ast|\hat{\mathcal{U}}|\psi_{s0};\psi_i^\ast\rangle\big|^2,
\end{align}
which is the duality. This method cannot determine $\arg C_U$. Also, the \emph{P}-function is often highly singular and might cause some problems.

From our analysis, the transfer matrix of the PTR setup can be seen in the Bogoliubov transformation of the nonlinear system whose causal order actually follows the transfer matrix. We believe this formalism can be further explored for a better understanding of quantum nonlinear optical processes.

\subsection{Discussion on the duality for single-mode squeezing}

For realistic linear systems with cavities, the forward- and backward-propagating lights have the same wavelength, so $\boldsymbol{L}_i=\boldsymbol{L}_s$ for a single linear lossless system (in this subsection, \emph{s} and \emph{i} only stand for forward and backward directions). One can find that $\boldsymbol{T}_{ii}=\boldsymbol{T}_{ss}^\ast$ and $\boldsymbol{T}_{is}=\boldsymbol{T}_{si}^\ast$ \cite{verifyt}. Then, with Eq.\ \eqref{symproperties}, the Bloch-Messiah reduction of $\boldsymbol{T}$ is possible \cite{PhysRevA.71.055801,matrix}, and the scattering matrix satisfies the reciprocity $\boldsymbol{\mathcal{U}}_{ii}=\boldsymbol{\mathcal{U}}_{ss}^\mathrm{T}$, $\boldsymbol{\mathcal{U}}_{si}=\boldsymbol{\mathcal{U}}_{si}^\mathrm{T}$, and $\boldsymbol{\mathcal{U}}_{is}=\boldsymbol{\mathcal{U}}_{is}^\mathrm{T}$.

By comparing the forms, $\boldsymbol{T}$ directly equals $\boldsymbol{F}$ (in the $\boldsymbol{\alpha},\boldsymbol{\alpha}^\ast$ basis) of a nonlinear system where realistic BSs are replaced by single-mode squeezing processes. Letting $\bar{\mathbf{a}}=(\boldsymbol{\alpha}^\mathrm{T},\boldsymbol{\alpha}^\dagger)^\mathrm{T}$ and $\mathbf{b}=(\boldsymbol{\beta}^\mathrm{T},\boldsymbol{\beta}^\dagger)^\mathrm{T}$, $Q(\boldsymbol{\beta})$ of the multimode coherent state $\boldsymbol{\alpha}$ after $\boldsymbol{F}$ still has the form of Eq.\ \eqref{gaussq}, while $\boldsymbol{\sigma}'_Q=(\boldsymbol{T}\boldsymbol{T}^\dagger+\boldsymbol{I}_2)/2$,
\begin{equation}\label{ncsms}
    \sqrt{\det(\boldsymbol{\sigma}'_Q)^{-1}}=|\det\boldsymbol{V}|=|\det\boldsymbol{T}_{ii}^{-1}|=|\det\boldsymbol{\mathcal{U}}_{ii}|,
\end{equation}
and the exponent
\begin{equation}\label{exposms}
    -\frac{1}{2}\left|\boldsymbol{V}(\mathbf{b}-\boldsymbol{T}\bar{\mathbf{a}})\right|^2=
    -\left|\begin{pmatrix}\boldsymbol{\beta}/\sqrt{2}\\\boldsymbol{\alpha}^\ast/\sqrt{2}\end{pmatrix}-\boldsymbol{\mathcal{U}}\begin{pmatrix}\boldsymbol{\alpha}/\sqrt{2}\\\boldsymbol{\beta}^\ast/\sqrt{2}\end{pmatrix}\right|^2
\end{equation}
is the same as $Q(\boldsymbol{\beta}/\sqrt{2},\boldsymbol{\alpha}^\ast/\sqrt{2})$ of the initial multimode coherent state $\boldsymbol{\alpha}/\sqrt{2}$ and $\boldsymbol{\beta}^\ast/\sqrt{2}$ (as the left and right inputs) after the corresponding linear system $\boldsymbol{\mathcal{U}}$. So, we have the duality for coherent input and postselected states
\begin{equation}\label{cohdualitysms}
    \big|\langle\boldsymbol{\beta}|\hat{U}|\boldsymbol{\alpha}\rangle\big|^2
    =|\det\boldsymbol{\mathcal{U}}_{ii}|\left|\left\langle\frac{\boldsymbol{\beta}}{\sqrt{2}};\frac{\boldsymbol{\alpha}^\ast}{\sqrt{2}}\right|\hat{\mathcal{U}}\left|\frac{\boldsymbol{\alpha}}{\sqrt{2}};\frac{\boldsymbol{\beta}^\ast}{\sqrt{2}}\right\rangle\right|^2.
\end{equation}
For example, considering the single-mode squeezing $\exp[r(\hat{a}^{\dagger2}-\hat{a}^2)/2]$ acting on the vacuum state, $\alpha=0$ and
\begin{equation}
    \boldsymbol{F}=\begin{pmatrix}\cosh r&\sinh r\\\sinh r&\cosh r\end{pmatrix},\;
    \boldsymbol{\mathcal{U}}=\begin{pmatrix}T&R\\-R&T\end{pmatrix}.
\end{equation}
From Eq.\ \eqref{ncsms} and the right-hand side of Eq.\ \eqref{exposms}, the \emph{Q}-function is calculated as $T\exp(-|\beta|^2+R\operatorname{Re}\beta^2)/\pi$, which is indeed for a squeezed vacuum state.

Because of the $1/\sqrt{2}$ factor and the repeated appearances of $\boldsymbol{\alpha}$ and $\boldsymbol{\beta}$ in Eq.\ \eqref{cohdualitysms}, a duality for arbitrary states may not be easily obtained.

\section{Four-photon advanced-wave picture}\label{fourawp}

Here, we describe the four-photon AWP for nondegenerate SPDC which suits for these two cases: (a) $|C_U|\approx1$ so that the probability of generating three or more pairs of photons is negligible (this requirement of $|C_U|$ is weaker than the two-photon AWP), and the postselection probability can be the fourfold (at four different paths, otherwise additional BSs should be used) coincidence counting probability; (b) photon number-resolving detectors are placed at all the output paths to ensure the \emph{s} paths and the \emph{i} paths have exactly two photons respectively. We mainly concern about the relative postselection amplitudes, so the NC is ignored in this section.

In case (a), without loss of generality, we let the four paths be $s1$, $s2$, $i1$, and $i2$. The four-fold coincidence counting probability
\begin{align}\label{fourprob}
    &\langle\mathrm{vac}|\hat{U}^\dagger\hat{a}_{s1}^\dagger\hat{a}_{s2}^\dagger\hat{a}_{i1}^\dagger\hat{a}_{i2}^\dagger\hat{a}_{s1}\hat{a}_{s2}\hat{a}_{i1}\hat{a}_{i2}\hat{U}|\mathrm{vac}\rangle\nonumber\\
    \approx\,&\langle\mathrm{vac}|\hat{U}^\dagger\hat{a}_{s1}^\dagger\hat{a}_{s2}^\dagger\hat{a}_{i1}^\dagger\hat{a}_{i2}^\dagger|\mathrm{vac}\rangle\langle\mathrm{vac}|\hat{a}_{s1}\hat{a}_{s2}\hat{a}_{i1}\hat{a}_{i2}\hat{U}|\mathrm{vac}\rangle\nonumber\\
    =\,&\big|\langle\mathbf{e}_{s1}+\mathbf{e}_{s2};\mathbf{e}_{i1}+\mathbf{e}_{i2}|\hat{U}|\mathrm{vac}\rangle\big|^2\nonumber\\
    \approx\,&\big|\langle\mathbf{e}_{s1}+\mathbf{e}_{s2};\mathbf{0}|\hat{\mathcal{U}}|\mathbf{0};\mathbf{e}_{i1}+\mathbf{e}_{i2}\rangle\big|^2,
\end{align}
so the probability amplitude is proportional to
\begin{align}
    &\langle\mathbf{e}_{s1}+\mathbf{e}_{s2};\mathbf{0}|\hat{\mathcal{U}}|\mathbf{0};\mathbf{e}_{i1}+\mathbf{e}_{i2}\rangle\nonumber\\
    =\,&\langle\mathrm{vac}|\hat{a}_{s1}\hat{a}_{s2}\hat{\mathcal{U}}\hat{a}_{i1}^\dagger\hat{\mathcal{U}}^\dagger\hat{\mathcal{U}}\hat{a}_{i2}^\dagger\hat{\mathcal{U}}^\dagger|\mathrm{vac}\rangle\nonumber\\
    =\,&\langle\mathrm{vac}|\hat{a}_{s1}\hat{a}_{s2}\Bigg(\sum_j\mathcal{U}_{sj,i1}\hat{a}_{sj}^\dagger\Bigg)\Bigg(\sum_j\mathcal{U}_{sj,i2}\hat{a}_{sj}^\dagger\Bigg)|\mathrm{vac}\rangle\nonumber\\
    =\,&\mathcal{U}_{s1,i1}\mathcal{U}_{s2,i2}+\mathcal{U}_{s2,i1}\mathcal{U}_{s1,i2}\nonumber\\
    =\,&\mathcal{U}_{s1,i1}(\mathcal{U}^\mathrm{T})_{i2,s2}+\mathcal{U}_{s2,i1}(\mathcal{U}^\mathrm{T})_{i2,s1},
\end{align}
which can be obtained from this classical scenario (except for the presence of hypothetical BSs): the $s1$ and $s2$ output paths are connected by a fiber; letting an \emph{i} light of unit amplitude emit from the $i1$ output path, it passes through the PTR setup with hypothetical BSs and becomes an \emph{s} light which passes through the fiber in both directions and goes back to the PTR setup to become another \emph{i} light; the probability amplitude is proportional to its amplitude at the $i2$ output path.

In case (b), if two photons are at the same path, we have
\begin{align}
    &\langle2\mathbf{e}_{s1};\mathbf{e}_{i1}+\mathbf{e}_{i2}|\hat{U}|\mathrm{vac}\rangle\nonumber\\
    \propto\,&\langle2\mathbf{e}_{s1};\mathbf{0}|\hat{\mathcal{U}}|\mathbf{0},\mathbf{e}_{i1}+\mathbf{e}_{i2}\rangle\nonumber\\
    =\,&\langle\mathrm{vac}|\frac{\hat{a}_{s1}^2}{\sqrt{2}}\Bigg(\sum_j\mathcal{U}_{sj,i1}\hat{a}_{sj}^\dagger\Bigg)\Bigg(\sum_j\mathcal{U}_{sj,i2}\hat{a}_{sj}^\dagger\Bigg)|\mathrm{vac}\rangle\nonumber\\
    =\,&\sqrt{2}\mathcal{U}_{s1,i1}(\mathcal{U}^\mathrm{T})_{i2,s1}
\end{align}
and
\begin{equation}
    \langle2\mathbf{e}_{s1};2\mathbf{e}_{i1}|\hat{U}|\mathrm{vac}\rangle\propto\mathcal{U}_{s1,i1}^2
    =\mathcal{U}_{s1,i1}(\mathcal{U}^\mathrm{T})_{i1,s1},
\end{equation}
which means the light at path $s1$ is directly reflected.

\begin{figure}[t]
\includegraphics[width=0.44\textwidth]{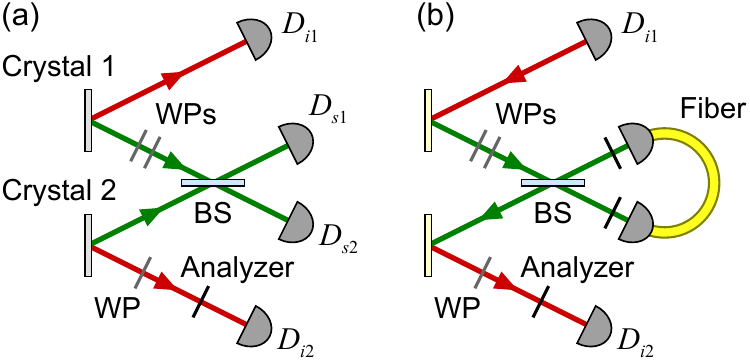}
\caption{\label{telefig}(a) Setup of the quantum teleportation experiment. The \emph{s} light from crystal 1 passes through wave plates (WPs) for polarization preparation. Crystal 2 produces polarization-entangled photon pairs. The \emph{s} photons from two crystals interfere at the BS. The \emph{i} photon from crystal 2 obtains the prepared polarization if detectors $D_{i1}$, $D_{s1}$, and $D_{s2}$ click. (b) Setup in the four-photon AWP. The fiber rotates the polarization by $90^\circ$ if the two polarization analyzers in front of it have orthogonal orientations.}
\end{figure}

Then, we discuss the quantum teleportation experiment \cite{Bouwmeester1997}, as shown in Fig.\ \ref{telefig}(a). Two low-gain nondegenerate SPDC crystals are used. Crystal 1 provides the photon (let it be \emph{s}) whose polarization is to be prepared by wave plates, and its \emph{i} partner only heralds its presence using the detector $D_{i1}$, so this SPDC crystal can be of any type, and we let it be type-I producing horizontally ($H$) polarized biphotons, whose AWP model is simply weakly reflecting $H$ lights. Crystal 2 (type-II) produces polarization-entangled biphotons $(|H\rangle_s|V\rangle_i-|V\rangle_s|H\rangle_i)/\sqrt{2}$ ($V$ for the vertical polarization), whose AWP model is rotating the polarization by $90^\circ$ during reflection (unchanged if the polarization is circular). The \emph{s} lights from the two crystals interfere at a $50:50$ BS, and detectors $D_{s1}$ and $D_{s2}$ are placed at its two output ports. As detectors are insensitive to the polarization, the detection probability should be the sum (incoherent superposition) of these four scenarios: two polarization analyzers with $(H,H)$, $(V,V)$, $(H,V)$, and $(V,H)$ (other orthogonal linear polarization pairs can be used; the fiber rotates the polarization by $90^\circ$ in the last two scenarios) are placed in front of them. The polarization of the \emph{i} light from crystal 2 is measured using a quarter-wave plate, a polarization analyzer, and detector $D_{i2}$. Letting the prepared polarization be $c_H|H\rangle+c_V|V\rangle$, the teleportation phenomenon is that the polarization of the light toward $D_{i2}$ is the same as the prepared one (up to a unitary transform) if the other three detectors click.

In the AWP setup as shown in Fig.\ \ref{telefig}(b), an \emph{i} light emits from $D_{i1}$, is reflected by crystal 1, and passes through the wave plates to prepare the polarization. Then, we consider the four scenarios. For $(H,H)$ and $(V,V)$, the transmitted light goes back to crystal 1 and does not contribute to coincidence counting; for $(H,V)$ and $(V,H)$, the polarization of the light heading toward crystal 2 is $-c_H|V\rangle+c_V|H\rangle$ and $c_H|V\rangle-c_V|H\rangle$ respectively, which means it has been rotated by $90^\circ$. After being reflected by crystal 2, its polarization becomes the prepared one again. So, the polarization has been teleported in the quantum setup.

This four-photon AWP avoids the HOM-like two-photon interference in a linear setup, which is present in the PTR theory of the main text. Besides quantum teleportation, it can also help analyze some quantum interference phenomena based on path identity \cite{Qian2023}. However, the four-photon term of degenerate SPDC and the six-photon (or more) term of SPDC involve more permutation terms and cannot be easily described by such a classical model.


\begin{thebibliography}{51}
\bibitem [{twm()}]{twm}%
  \BibitemOpen
  \href@noop {} {\bibinfo {title} {{It can also be a three-wave mixing process
  with a strong coherent light under the parametric
  approximation.}}}\BibitemShut {Stop}%
\bibitem [{svd()}]{svd}%
  \BibitemOpen
  \href@noop {} {\bibinfo {title} {{The singular values of a scattering matrix
  for a linear optical system cannot be greater than $1$. If their product (the
  modulus of the determinant) is $1$, they are all $1$ and the matrix is
  unitary.}}}\BibitemShut {Stop}%
\bibitem [{rea()}]{realcavity}%
  \BibitemOpen
  \href@noop {} {\bibinfo {title} {{The optical cavity can be seen in an
  optical parametric oscillator or amplifier, where a nonlinear crystal is
  inserted into it. Nevertheless, this theory is suitable for a well-calibrated
  setup which performs the PDC with almost no other influences on the optical
  field.}}}\BibitemShut {Stop}%
\bibitem [{pol()}]{polyfilter}%
  \BibitemOpen
  \href@noop {} {\bibinfo {title} {{In practice, there might be PDC processes
  outside $D_s$ or $D_i$. However, perfect interference filters can be placed
  to totally absorb lights with unwanted frequencies which are isolated from
  our considered ones. If the pump beam is polychromatic, the filtering leads
  to undetected paths coupled to detected ones.}}}\BibitemShut {Stop}%
\bibitem [{ver()}]{verifyt}%
  \BibitemOpen
  \href@noop {} {\bibinfo {title} {{These are proved by verifying them for the
  elementary cases, one BS and one linear system pair, and then proving if two
  transfer matrices do, their product also does.}}}\BibitemShut {Stop}%
\end{thebibliography}
\end{document}